\documentclass[prd,aps,twocolumn,showpacs,showkeys,nofootinbib,floatfix]{revtex4-2}

\usepackage{amsmath,amssymb}
\usepackage{graphicx}
\usepackage[colorlinks=true,linkcolor=blue,citecolor=blue,urlcolor=violet]{hyperref}
\usepackage[normalem]{ulem}
\usepackage{xcolor}
\usepackage{cancel}
\usepackage{xcolor}
\graphicspath{ {./Figures/} }

\usepackage[caption=false]{subfig}

\newcommand{\redsout}[1]{%
  \bgroup
  \markoverwith{\textcolor{red}{\rule[0.5ex]{2pt}{1.2pt}}}%
  \ULon{#1}%
}

\begin{document}

\title{
Bayesian Constraints on Bosonic Dark Matter in Neutron Stars within a Covariant Density-Functional Equation-of-State Ensemble
}

\author{Davood Rafiei Karkevandi$^{1}$}
\email{davood.rafiei@uwr.edu.pl}

\author{Alexander~Ayriyan$^{1,2}$}
\email{alexander.ayriyan@gmail.com}

\author{Armen Sedrakian$^{1,3}$}
\email{armen.sedrakian@uwr.edu.pl}

\affiliation{$^1$Institute of Theoretical Physics, University of Wroc\l{}aw, 50-204 Wroc\l{}aw, Poland}
\affiliation{$^2$Alikhanyan National Science Laboratory (Yerevan Institute of Physics), 0036 Yerevan, Armenia}
\affiliation{$^3$Frankfurt Institute for Advanced Studies, D-60438 Frankfurt am Main, Germany}

\date{\today}

\begin{abstract}
 We investigate neutron stars admixed with repulsively self-interacting
bosonic dark matter (DM) using current multi-messenger observations.
The dark component is modeled as a complex scalar field coupled to
baryonic matter only through gravity, allowing both dark-core and
dark-halo configurations. To quantify baryonic equation-of-state (EOS)
uncertainties, we combine a three-dimensional scan over
$(m_\chi,\lambda,F_\chi)$ with 27 DDME2-based covariant
density-functional EOSs spanning
$L_{\rm sym}=30,50,70$ MeV and
$-600\le Q_{\rm sat}\le 1000$ MeV. The models are confronted in a
Bayesian framework with NICER mass--radius measurements from the
Amsterdam and Maryland/Illinois analyses and with the GW170817
tidal-deformability constraint.
After Bayesian evidence weighting over the EOS ensemble, the
EOS-relevant dark-sector scale
$\mu_\chi\equiv m_\chi/\lambda^{1/4}$ is localized around
$\mu_\chi\simeq250$ MeV for both NICER data sets. In the original
$(m_\chi,\lambda)$ parameterization, the marginalized boson-mass
posterior peaks at $m_\chi\simeq300$--$330$ MeV, while $\lambda$
remains broadly distributed. Since the bosonic EOS depends on
$m_\chi$ and $\lambda$ only through $\mu_\chi$, their separate
marginals are conditional on the adopted independent priors. The
inferred DM fraction is more data-set dependent, with posterior modes
near $F_\chi\simeq4\%$ for Maryland/Illinois and
$F_\chi\simeq12\%$ for Amsterdam.
Allowing for bosonic DM also shifts the inferred high-density nuclear
parameter $Q_{\rm sat}$ from negative toward intermediate and positive
values, whereas the response of $L_{\rm sym}$ is weaker. This indicates
a model-dependent trade-off between the compactifying effect of the
dark component and the high-density stiffness of baryonic matter.
These constraints remain conditional on the selected EOS family and on
the assumption of a common effective $F_\chi$ for all sources.

\end{abstract}


\maketitle

\section{Introduction}

The nature of dark matter (DM) remains one of the central open problems in modern physics \cite{Cirelli:2024ssz}. Neutron stars (NSs) provide a promising environment for testing DM scenarios because of their extreme densities and deep gravitational potentials. If DM accumulates in their interiors, it can modify the stellar structure and leave observable signatures \cite{Bramante:2023djs,Grippa:2024ach,Brax:2026cmh}.

Several formation and accumulation channels have been proposed for DM admixed NSs \cite{Ellis:2018bkr,Gresham:2018rqo,Kouvaris:2010jy,Ciarcelluti:2010ji}. In general, the resulting DM content depends on the DM particle properties, the DM--baryon interaction strength, the stellar properties and evolutionary history, and the ambient DM density and velocity distribution, which vary with the Galactic environment \cite{Karkevandi:2021ygv,Ivanytskyi:2019wxd,Takatsy:2025nnw,Liang:2023nvo,Bell:2020obw,DelPopolo:2019nng}. DM may be captured during the protostellar and main-sequence stages \cite{Gupta:2025ygk,Robles:2025dbd} and may continue to accumulate during the subsequent evolution of the star \cite{Ebadi:2025umm,Sarkar:2026fge,Parmar:2025bux}.  During the short-lived, hot, and dense proto-NS stage, dark particles may be produced rapidly through model-dependent thermal processes, with a gravitationally bound component potentially surviving as the star cools~\cite{Nelson:2018xtr, Pospelov:2026dnb}. In a mature, equilibrated NS, additional dark particles may be produced through neutron-to-dark-sector conversion or other density-induced in-medium reactions \cite{Baym:2018ljz,Harris:2025isp,McKeen:2021jbh}, while further capture from the surrounding DM halo may continue over the stellar lifetime \cite{Liu:2025qco,Bell:2020jou,Sandin:2008db}. Other possibilities include formation in DM-rich environments or minihalos \cite{Arvikar:2026kci}, baryonic accretion onto a pre-existing dark core \cite{Kamenetskaia:2022lbf}, accretion of DM from dark compact objects and mergers involving dark and baryonic compact objects \cite{Dietrich:2018jov,Clough:2018exo}. These scenarios may produce DM fractions ranging from very small values to percent-level or larger admixtures.

The impact of DM on compact stars depends strongly on its assumed interaction with baryonic matter (BM). In single-fluid descriptions, the dark component may be produced inside the star \cite{Shahrbaf:2024gdm,Shahrbaf:2022upc,Das:2025pjl,Heeck:2026hgy} or accumulated through capture and accretion, and is assumed to remain locally mixed with BM through non-gravitational interactions \cite{Sedaghat:2025dzt,Issifu:2026diw,Hajkarim:2024ecp,Sen:2024yim,Klangburam:2025rcb,Kumar:2026hoq}. Its presence modifies the microscopic properties of dense matter, generally softens the effective equation-of-state (EOS) , and affects the observable properties of the star. In two-fluid models, by contrast, BM and DM interact only through gravity \cite{Shakeri:2022dwg,Sedaghat:2026okv,Pitz:2024xvh,Thakur:2023aqm,Das:2025fyf,Dengler:2025ntz,Biesdorf:2024dor,Lourenco:2025ugk,Routaray:2024fcq,karan:2025kel}. The dark component may then form either a compact core or an extended halo \cite{RafieiKarkevandi:2021hcc}. Dark cores usually reduce the visible baryonic radius and the maximum supported mass, whereas extended dark halos can increase the outer gravitational radius, enhance the tidal response, and raise the maximum supported mass \cite{Karkevandi:2024vov}.

DM may affect compact stars in several evolutionary stages and through several observables.  Beyond modifying the masses, radii, and tidal deformabilities of compact stars, an admixed dark component may affect their rotational properties \cite{Cronin:2023xzc,Cipriani:2025tga,Cipriani:2025wem,Konstantinou:2024ynd}, oscillation spectra \cite{Shahrbaf:2025hsw,Zhang:2026rkg,Routaray:2025gbq}, thermal evolution \cite{Issifu:2025jac,AngelesPerez-Garcia:2022qzs}, cooling \cite{Sedrakian:2015krq,Sedrakian:2018kdm,Avila:2023rzj,Zhou:2025dmy} and heating rates \cite{Baryakhtar:2017dbj,Kouvaris:2007ay,Bell:2018pkk}, neutrino emission \cite{Issifu:2024htq,Kumar:2022amh}, X-ray pulse profiles \cite{Shakeri:2022dwg,Shawqi:2024jmk} and gravitational lensing \cite{Kumaran:2026zbt}. It may also leave signatures in the inspiral, merger, and post-merger dynamics of compact binaries and in the associated gravitational-wave signals \cite{Kunert:2026hmx,Koehn:2024gal,Bauswein:2020kor,Mahapatra:2026utt,Giangrandi:2025rko,Srikanth:2025lic}.

Among the various DM candidates, bosonic DM is particularly well motivated in a wide range of astrophysical environments, including compact objects such as NSs \cite{Karkevandi:2021ygv,Shahrbaf:2023uxy,Rather:2026aed,Lam:2025for,Sedaghat:2026xit,Fakhry:2026cmi,Giangrandi:2022wht,Diedrichs:2023trk}. Although the Standard Model contains several elementary bosons, its massive bosons are unstable, while its stable bosons are massless. Stable massive bosons predicted by physics beyond the Standard Model are therefore well-motivated DM candidates. Self-gravitating bosonic configurations have also been widely studied as boson stars \cite{Liebling:2012fv}, in which quantum pressure, supplemented by repulsive self-interactions, can support compact objects with astrophysically relevant masses and radii \cite{Pitz:2023ejc,RafieiKarkevandi:2021hcc,Chavanis:2022fvh,Koliogiannis:2024szd}.  Such horizonless objects may mimic some black-hole observables while producing distinctive signatures through gravitational lensing, accretion, oscillations, and gravitational waves \cite{Evstafyeva:2026juq,Brihaye:2025dlq,Shirke:2025ust,Collier:2022cpr}. Motivated by these considerations, in this work we adopt a bosonic-DM model described by a complex scalar field with a repulsive quartic self-interaction, characterized by the boson mass $m_\chi$ and coupling $\lambda$. When admixed with BM in a two-fluid NS, the stellar structure also depends on the DM fraction inside the star, $F_\chi$.

A major challenge in studies of DM-admixed NSs is that the inferred dark-sector constraints can depend strongly on the adopted BM EOS. Changes in the stiffness of the baryonic EOS can produce effects similar to those caused by DM, making it difficult to distinguish between them, especially at supranuclear densities where the BM EOS is still uncertain. To reduce the dependence of our results on any single baryonic model, we consider an ensemble of DDME2-based covariant density functional (CDF) EOSs in which the symmetry-energy slope $L_{\rm sym}$ and the skewness parameter $Q_{\rm sat}$ are varied systematically. This allows the baryonic-EOS uncertainty to be quantified and propagated into the inferred DM parameters. The parameter $L_{\rm sym}$ mainly affects NS radii and tidal deformabilities, whereas $Q_{\rm sat}$ controls the high-density stiffness and is particularly important for massive stars.  Still,  the EOSs used in this work constitute a useful controlled DDME2-based ensemble, but they do not represent general nuclear-EOS uncertainty. Only two selected nuclear parameters are varied, on a coarse grid, while the functional form and the remaining saturation properties are fixed.

Bayesian methods have increasingly been applied to both fermionic and bosonic DM-admixed NSs, considering different baryonic-EOS treatments \cite{Arvikar:2025dwl,Arvikar:2025hej,Rutherford:2024uix,Rutherford:2022xeb,Liu:2025cwy,Santos:2025xep,Sen:2026qfx,Grippa:2024sfu,Kunert:2026hmx,Thakur:2024btu,Rutherford:2026dxk,Shahrbaf:2024gdm}. In this work, we perform a Bayesian analysis of self-interacting bosonic DM admixed with NS matter using 27 CDF EOSs with systematic variations of $L_{\rm sym}$ and $Q_{\rm sat}$, together with a three-dimensional scan over $(m_\chi,\lambda,F_\chi)$. The resulting stellar models are tested against NICER mass–radius measurements from the Amsterdam and Maryland/Illinois analyses and the GW170817 tidal deformability bound. The fixed-EOS posteriors are then combined through Bayesian evidence weighting, yielding model-averaged constraints on the bosonic-DM parameters while consistently accounting for the uncertainty across the baryonic-EOS ensemble. The resulting EOS likelihoods and marginalized posteriors for $L_{\rm sym}$ and $Q_{\rm sat}$ show how the inclusion of DM shifts the preferred nuclear-parameter region. Finally, we compare the Bayesian evidence for DM-admixed and purely baryonic models. We note that, for halo configurations, the published NICER mass–radius constraints are applicable only approximately, since the pulse-profile analyses underlying these constraints assume a vacuum exterior and do not account for the gravitational effect of an extended DM halo outside the baryonic surface.  However, extended halo configurations are strongly suppressed over much of the relevant parameter space by the GW170817 tidal-deformability constraint, reducing the impact of this approximation on the final posterior.  The approximate nature of the NICER constraint for halo configurations is thus unlikely to have a material effect on our results.

In the present analysis, we assume a common DM fraction $F_\chi$ for all NSs entering the observational data set, while the particle parameters $m_\chi$ and $\lambda$ are likewise taken to be universal. This simplifying assumption, which has also been adopted in previous Bayesian studies of DM-admixed NSs, allows the multi-messenger constraints to be incorporated within a tractable three-parameter Bayesian framework and should be regarded as an effective population-level approximation rather than a detailed model of DM accumulation in individual stars. In reality, the DM fraction may depend on the stellar formation history, age, progenitor properties, ambient DM environment, capture or conversion mechanism, and possibly the stellar mass; consequently, different NICER sources and even the two components of GW170817 need not contain identical dark fractions. The inferred constraints on $F_\chi$, and through parameter correlations also on $m_\chi$, are therefore conditional on the universal-$F_\chi$ assumption and should not be interpreted as applying directly to a completely general population of DM-admixed NSs. A more comprehensive treatment would introduce a source-dependent fraction $F_{\chi,i}$ as a nuisance parameter for each object, or infer a hierarchical population distribution of dark fractions.

This paper is structured as follows. Section~\ref{sec:eos} introduces the CDF EOSs used to describe the baryonic component, while Sec.~\ref{sec:dark_matter} presents the self-interacting bosonic DM model. In Sec.~\ref{sec:admixed_NS}, we summarize the two-fluid stellar-structure equations and define the relevant masses, radii, and DM fraction. Sections~\ref{sec:nodm_baseline} and \ref{sec:bosonic_DM_admixed_NS} present, respectively, the purely baryonic reference configurations and the sensitivity of the stellar sequences to the dark-sector parameters.  In Sec.~\ref{sec:DM_Core-Halo-distribution}, we analyze the transition between dark-core and dark-halo configurations and examine the associated changes in the maximum stellar mass. Section~\ref{sec:Bayesian_analysis} describes the Bayesian framework, the observational constraints, and the posterior-weighted modifications produced by the bosonic admixture. Section~\ref{sec:Results} presents the evidence-weighted constraints on the dark-sector parameters, the inferred nuclear EOS parameters, and the Bayesian comparison between the DM-admixed and purely baryonic model families. The broader physical implications and limitations of the analysis and our main conclusions are summarized in Sec.~\ref{sec:conclusions}.

\section{Covariant Density Functional Equations of State} \label{sec:eos}

The baryonic component is described by the covariant density functional (CDF) EOSs of Ref.~\cite{Li:2023bid}. This framework provides a relativistic description of dense nuclear matter with density-dependent meson-nucleon couplings and allows controlled variations of poorly constrained nuclear matter parameters. It is therefore well suited for propagating baryonic EOS uncertainty into the inferred bosonic-DM constraints.

We use the DDME2-based CDF family and select 27 purely nucleonic EOSs from the larger set of Ref.~\cite{Li:2023bid}. The grid points in the plane $L_{\rm sym}$--$Q_{\rm sat} $ are defined by the following choice:
%-----------------------
\begin{align}
L_{\rm sym} &= 30,\;50,\;70~{\rm MeV}, \\
Q_{\rm sat} &\in [-600,1000]~{\rm MeV},\quad \Delta Q_{\rm sat} = 200~{\rm MeV}
\end{align}
%-----------------------
where $\Delta Q_{\rm sat} $ is the step between adjacent grid points on the $Q_{\rm sat}$ axis. 
Here $L_{\rm sym}$ controls the density dependence of the symmetry energy and affects radii and tidal deformabilities, while $Q_{\rm sat}$ controls the high-density behavior of symmetric nuclear matter and strongly affects the stiffness of massive-star configurations.

For each EOS, cold catalyzed matter is constructed by imposing beta equilibrium and charge neutrality. The resulting EOSs are used as the baryonic input in the two-fluid stellar-structure calculations. Applying the same bosonic-DM scan to all 27 EOSs allows us to quantify how the uncertainty in $(L_{\rm sym},Q_{\rm sat})$ propagates into the inferred dark-sector parameters. In the Bayesian analysis, each EOS is later weighted by its evidence, so that EOSs more compatible with the same multi-messenger data contribute more significantly to the final model-averaged posterior.

\section{Bosonic Dark Matter Model} \label{sec:dark_matter}

We describe the dark component as a complex scalar field with a repulsive quartic self-interaction~\cite{Colpi:1986ye}. The Lagrangian density is
\begin{equation}
\mathcal{L}
=
\frac{1}{2}\partial_\mu\phi^*\partial^\mu\phi
-\frac{m_\chi^2}{2}\phi^*\phi
-\frac{\lambda}{4}(\phi^*\phi)^2 ,
\end{equation}
where $m_\chi$ is the boson mass and $\lambda$ is the dimensionless self-interaction coupling. The quartic term is repulsive and provides additional pressure support for the dark fluid.

In the strong-coupling/mean-field regime, the bosonic matter is described by the effective perfect-fluid EOS~\cite{Karkevandi:2021ygv}
\begin{equation}\label{eq:bosonic_eos}
P_\chi
=
\frac{m_{\chi}^{4}}{9\lambda}
\left(
\sqrt{1+\frac{3\lambda}{m_{\chi}^{4}}\rho_\chi}
-1
\right)^{2},
\end{equation}
where $P_\chi$ and $\rho_\chi$ are the pressure and energy density of the dark component. Smaller $m_\chi$ and larger $\lambda$ generally make the dark EOS stiffer, while heavier bosons or weaker self-interactions favor more compact dark configurations.

Equation~\eqref{eq:bosonic_eos} depends on $m_\chi$ and $\lambda$ through the combination $\lambda/m_\chi^4$. Consequently, within the effective perfect-fluid description adopted here, stellar-structure observables do not independently determine the two parameters: different pairs $(m_\chi,\lambda)$ with the same value of $\lambda/m_\chi^4$ generate the same dark-sector EOS. We nevertheless retain $m_\chi$ and $\lambda$ as separate sampling parameters because they are the microscopic parameters entering the underlying scalar-field Lagrangian and because the prior ranges and perturbativity condition are naturally formulated in terms of them. The corresponding one-dimensional marginalized constraints should therefore be understood as conditional on the adopted joint prior in $(m_\chi,\lambda)$.

This EOS is used in the two-fluid stellar-structure calculation, where BM and DM interact only through gravity. Depending on the values of the triplet $(m_\chi,\lambda,F_\chi)$, the dark component can appear either as a compact core or as an extended halo. In this study we scan the following range of this triplet of parameters of DM:
\begin{align}
m_\chi &= 100\text{--}1000~{\rm MeV},~\label{eq:mass_range} \\
\lambda &= 0.1\text{--}12.5,~\label{eq:lambda_range} \\
F_\chi &= 1\text{--}20\%~\label{eq:fraction_range} .
\end{align}
The upper limit $\lambda_{\rm max}=12.5$ is chosen just below $4\pi$, which we adopt as a conventional heuristic perturbativity cutoff rather than as a model-independent perturbative-unitarity bound. The strong-self-interaction condition $\lambda \gg 4\pi(m_\chi/M_{\rm Pl})^2$, with $M_{\rm Pl}$ being the Planck mass, underlying the effective fluid description~\cite{Colpi:1986ye} is comfortably satisfied throughout the parameter range considered here.
This range covers both core-like and halo-like configurations. The three-dimensional space of parameter triplet is discretized to yield 22 values of $\lambda$, 55 values of $m_\chi$, and 20 values of $F_\chi$. Therefore, for each fixed $(L_{\rm sym},Q_{\rm sat})$ pair, we calculate $22\times55\times20=24\,200$ dark-sector configurations, corresponding to a total of $27\times24\,200=653\,400$ configurations over the full nucleonic-EOS ensemble.

\section{Dark Matter Admixed Neutron Stars} \label{sec:admixed_NS}

We model the DM admixed NS  within the two-fluid Tolman-Oppenheimer-Volkoff (TOV) formalism~\cite{Karkevandi:2021ygv,Shakeri:2022dwg}. The baryonic and dark components are treated as separate perfect fluids that interact only through the common gravitational field. Each component satisfies its own hydrostatic-equilibrium equation, while both contribute to the total enclosed mass.

The two-fluid TOV equations are
\begin{align}
  \label{eq:TOV_1}
\frac{dp_{BM}}{dr}
&=
-(p_{BM}+\epsilon_{BM})
                     \frac{M+4\pi r^3 p}{r(r-2M)},\\
    \label{eq:TOV_2}
\frac{dp_{DM}}{dr}
&=
-(p_{DM}+\epsilon_{DM})
                     \frac{M+4\pi r^3 p}{r(r-2M)},\\
    \label{eq:TOV_3}
\frac{dM}{dr}
&=
4\pi r^2(\epsilon_{BM}+\epsilon_{DM}),
\end{align}
where
\begin{equation}
p=p_{BM}+p_{DM},
\end{equation}
and $M(r)=M_{BM}(r)+M_{DM}(r)$ is the total enclosed gravitational mass.

For a given pair of central pressures, the equations are integrated outward until each fluid pressure vanishes. This defines the baryonic radius $R_{BM}$ and dark radius $R_{DM}$. Configurations with $R_{DM}<R_{BM}$ contain a dark core, whereas those with $R_{DM}>R_{BM}$ contain an extended dark halo. The total gravitational mass is
\begin{equation}
M=M_{BM}(R_{BM})+M_{DM}(R_{DM}).
\end{equation}
For electromagnetic observables, we identify the visible stellar radius with $R_{BM}$, since photons are emitted from the baryonic surface. The dark fraction is defined as
\begin{equation}
F_\chi=\frac{M_{DM}(R_{DM})}{M}.
\end{equation}
Together with $m_\chi$ and $\lambda$, the DM fraction determines the structure of the admixed star. The two-fluid stellar sequences at fixed $F_\chi$ are constructed following Ref.~\cite{Karkevandi:2021ygv} by solving the two-fluid TOV equations \eqref{eq:TOV_1}-\eqref{eq:TOV_3}.
For each target $F_\chi$, the central pressures of the two fluids are adjusted so that the integrated masses satisfy $F_\chi=M_\chi/(M_B+M_\chi)$; the integration is continued until the pressure of each component vanishes, thereby determining its respective radius.

\section{Purely Baryonic Reference Models} \label{sec:nodm_baseline}

\begin{figure*}[thb] 
    %\centering 

    \includegraphics[width=0.49\textwidth] 
    {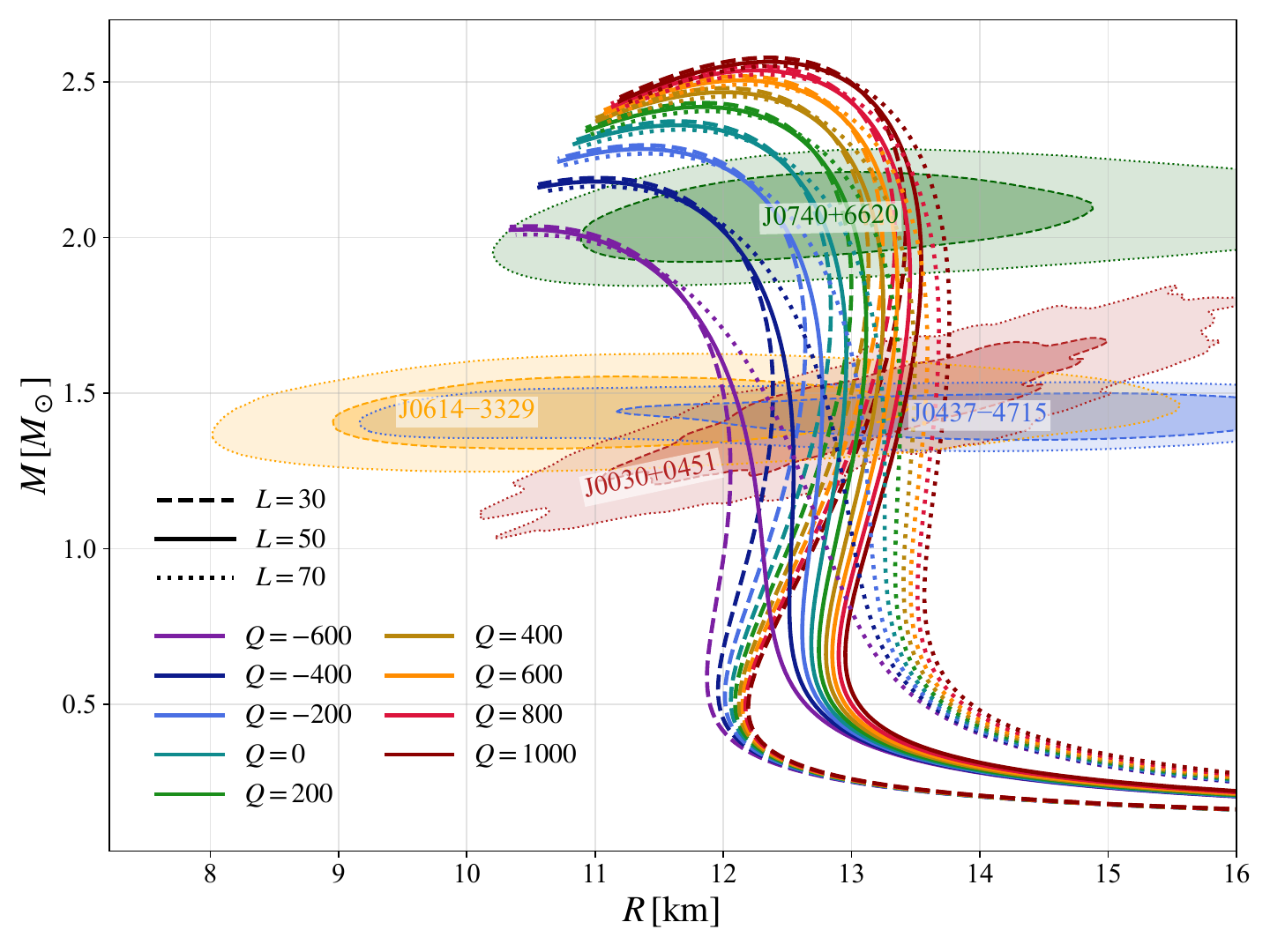}% 
    \hfill 
    \includegraphics[width=0.49\textwidth] 
    {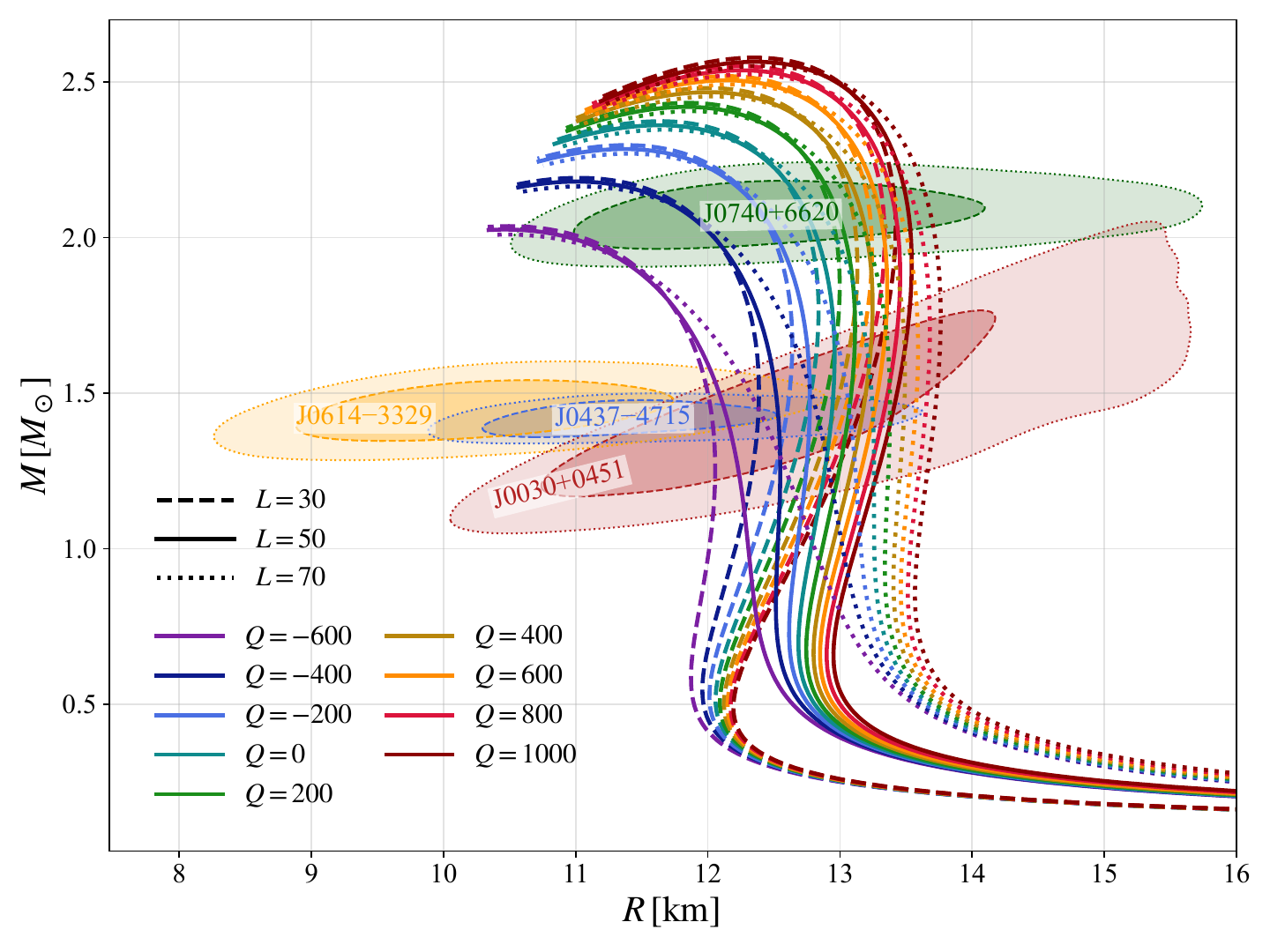}% 

    \vspace{-0.3em}
\caption{Purely baryonic NS sequences for the 27 CDF 
    EOSs used in this work. The line color denotes the 
    value of the high-density skewness parameter $Q_{\rm sat}$, while 
    the line style denotes the symmetry-energy slope $L_{\rm sym}$. 
    Left and right panels show the mass--radius relations compared with 
    the Maryland/Illinois and Amsterdam NICER mass--radius constraints, 
    respectively. These panels provide
    the no-DM reference against which the bosonic-DM-admixed results 
    are compared, see also~Ref.~\cite{Li:2023bid}.} 
    \label{fig:nodm_baseline} 
\end{figure*}
  
Before introducing the Bayesian inference of the bosonic-DM parameters, it is useful to examine the purely baryonic predictions of the selected CDF EOS ensemble, see also Ref.~\cite{Li:2023bid}. This provides the no-DM reference against which the DM-admixed configurations are compared. The 27 nucleonic EOSs span the $L_{\rm sym}$--$Q_{\rm sat}$ grid defined in Sec.~\ref{sec:eos} and therefore cover a broad range of radii, maximum masses, and tidal deformabilities.

Figure~\ref{fig:nodm_baseline} shows the mass--radius relations of the purely baryonic
sequences. The color of each curve represents the value of
$Q_{\rm sat}$, while the line style denotes $L_{\rm sym}$. The two
panels compare the same EOS ensemble with the Maryland/Illinois and
Amsterdam NICER mass--radius analyses, respectively. The comparison
illustrates that the two NICER data sets probe somewhat different
regions of the mass--radius plane, which later leads to different
posterior preferences for both the nuclear EOS parameters and the
bosonic-DM fraction.

\begin{figure*}[thb]
   \centering
    \includegraphics[width=0.49\textwidth] 
    {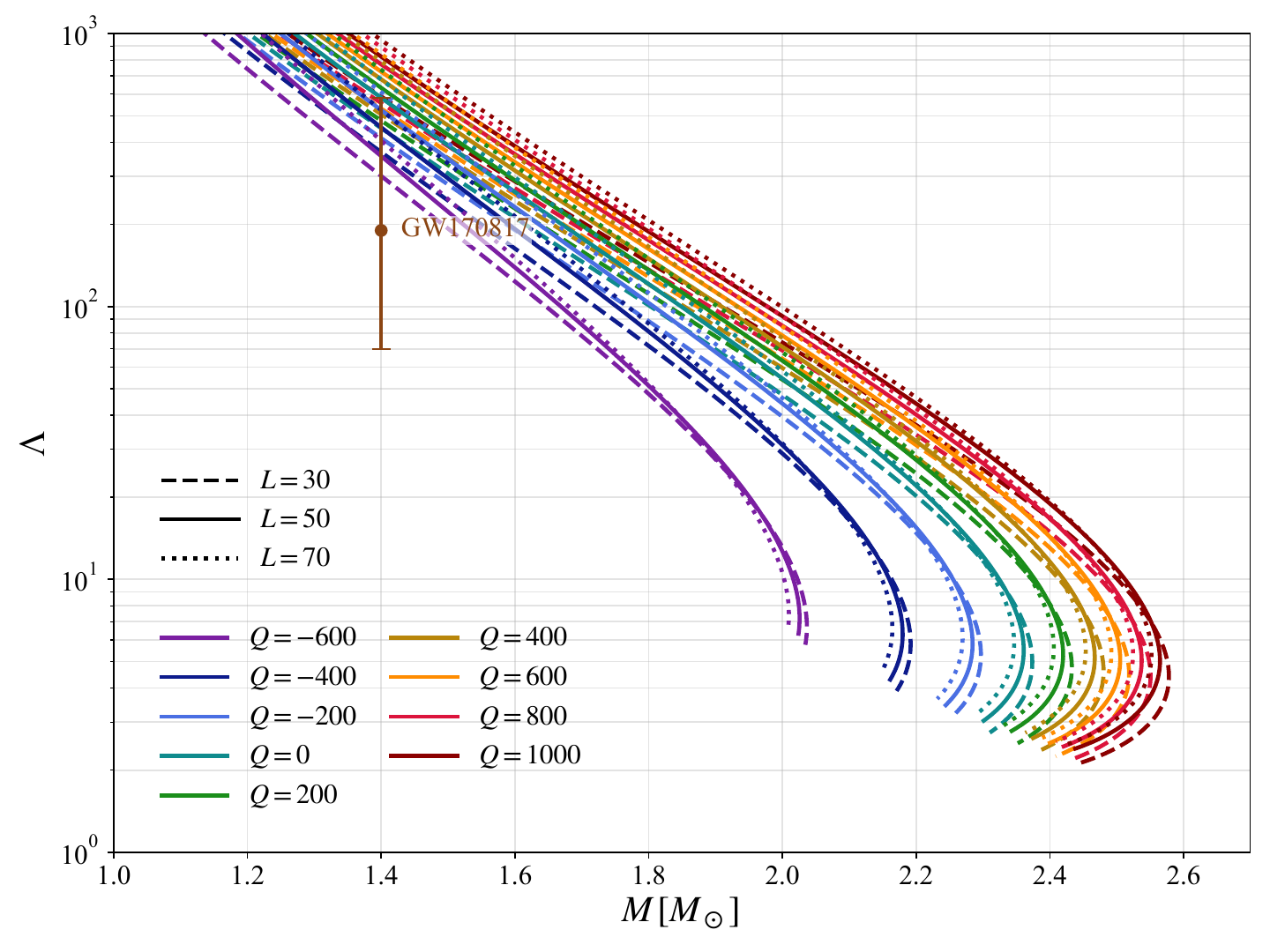}% 
    \centering
    \includegraphics[width=0.49\textwidth]
    {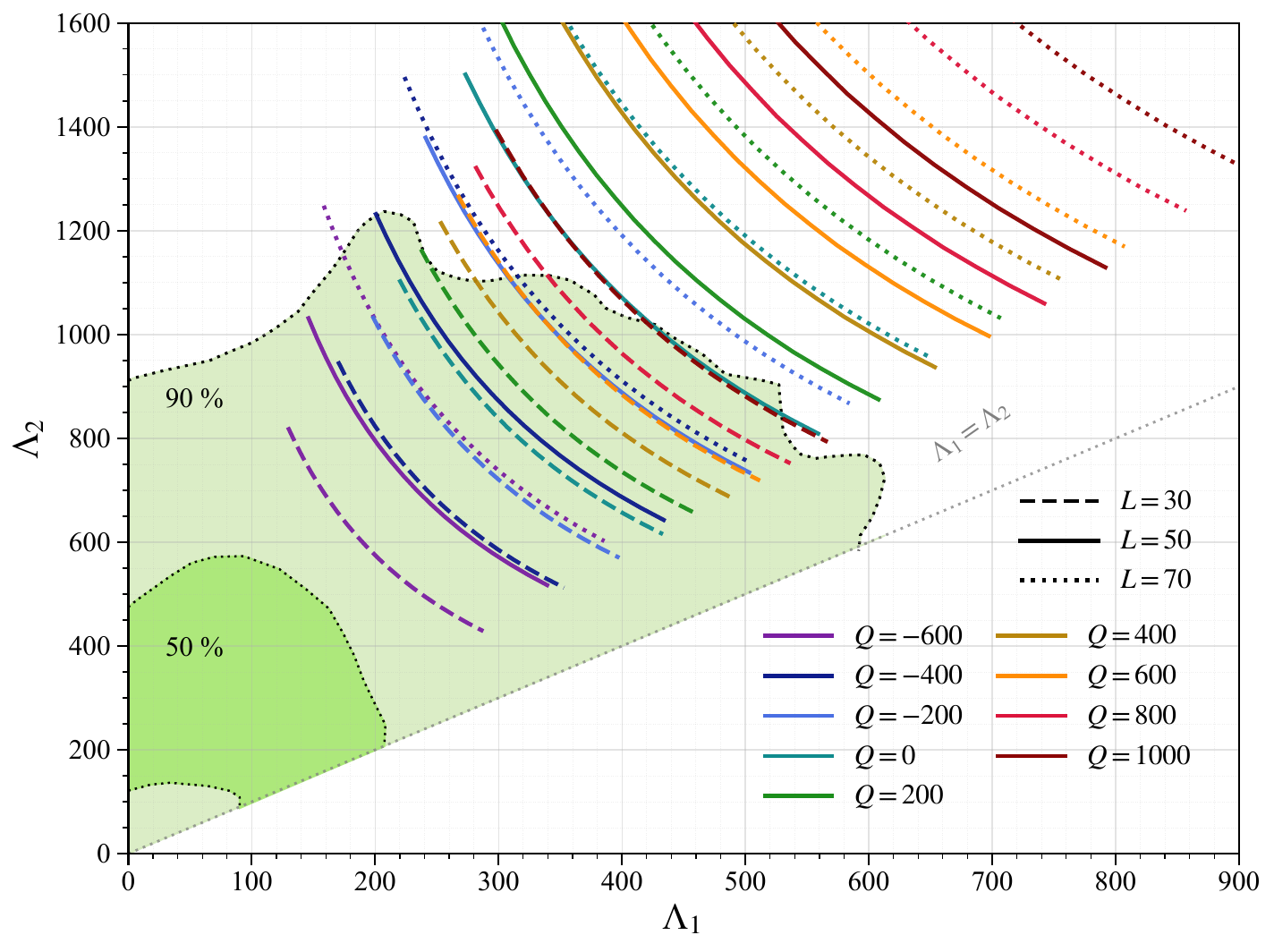}%
    \caption{
      Tidal deformabilities of the purely baryonic NS sequences
for the 27 CDF EOSs used in this work. The line color
denotes $Q_{\rm sat}$ and the line style denotes $L_{\rm sym}$. Left panel shows the dimensionless tidal deformability $\Lambda$ as a
function of gravitational mass; the marker at $M=1.4\,M_\odot$
provides a reference for the tidal response in the canonical-mass
range. Right panel shows the corresponding binary tidal-deformability
curves in the $\Lambda_1$--$\Lambda_2$ plane together with the
GW170817 credible regions. The figure provides the no-DM tidal
reference, see also Ref.~\cite{Li:2023bid},  used in the subsequent analysis.
  }
    \label{fig:nodm_lambda_mass}
\end{figure*}

Figure~\ref{fig:nodm_lambda_mass} summarizes the corresponding tidal-deformability behavior; it shows $\Lambda-M$ dependence and $\Lambda_1$--$\Lambda_2$ curves for the same EOS ensemble together with the GW170817 credible regions. These two complementary representations illustrate how the selected purely baryonic EOS ensemble spans the tidal response relevant for GW170817 before a dark component is introduced. As expected, the tidal deformability decreases rapidly with increasing stellar mass, reflecting the increasing compactness of heavier stars. At a fixed mass, stiffer EOSs generally predict larger values of $\Lambda$, while softer EOSs give smaller tidal deformabilities. The marker at $M=1.4\,M_\odot$ indicates the canonical-mass tidal constraint used as a reference in the analysis. 

This purely baryonic baseline is used in two ways in the following Bayesian analysis. First, it provides the reference likelihoods and evidences for the no-DM interpretation. Second, it allows us to identify how the inferred nuclear parameters change once the bosonic-DM component is introduced and marginalized over. In particular, comparing the no-DM and with-DM posteriors will show whether the preference for a given region of $(L_{\rm sym},Q_{\rm sat})$ is intrinsic to the baryonic EOS or can be partially absorbed by the presence of bosonic DM.

\section{Sensitivity to the dark matter parameters} \label{sec:bosonic_DM_admixed_NS}

\begin{figure*}
    \centering
        \includegraphics[width=0.32\linewidth]{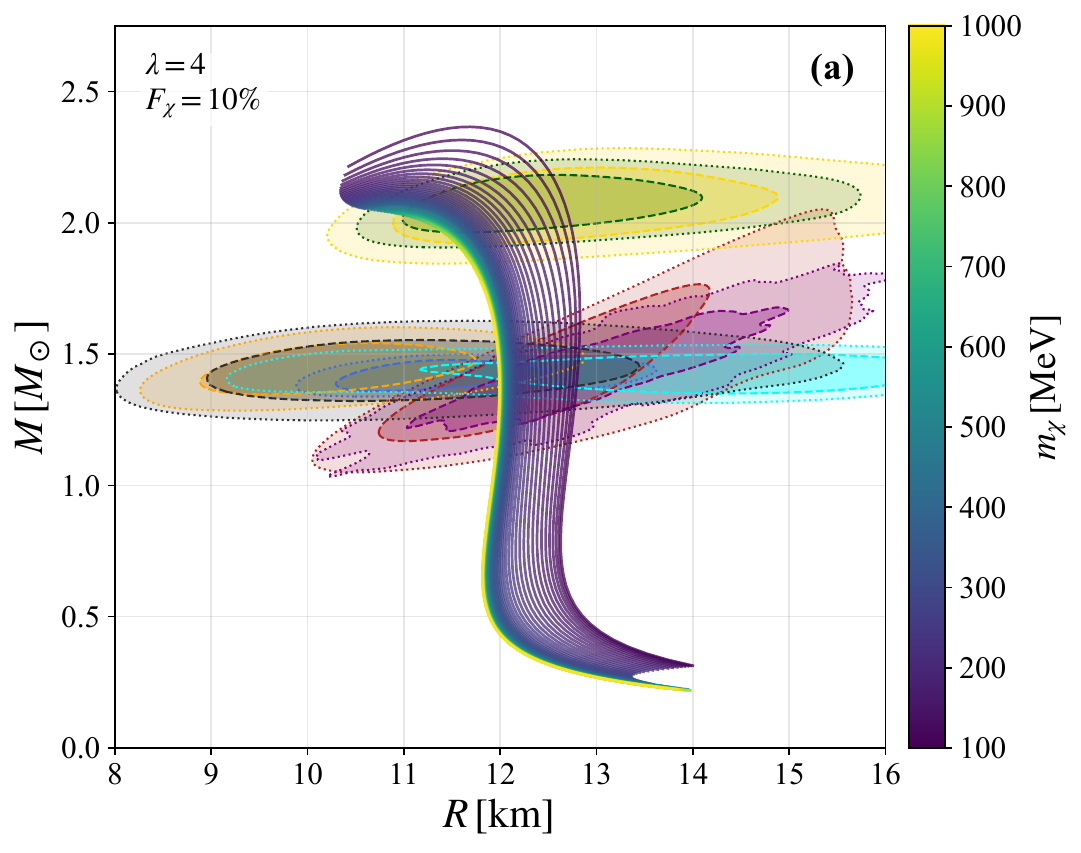}
        \label{fig:param_sensitivity_mchi}
    \hfill
        \includegraphics[width=0.32\linewidth]{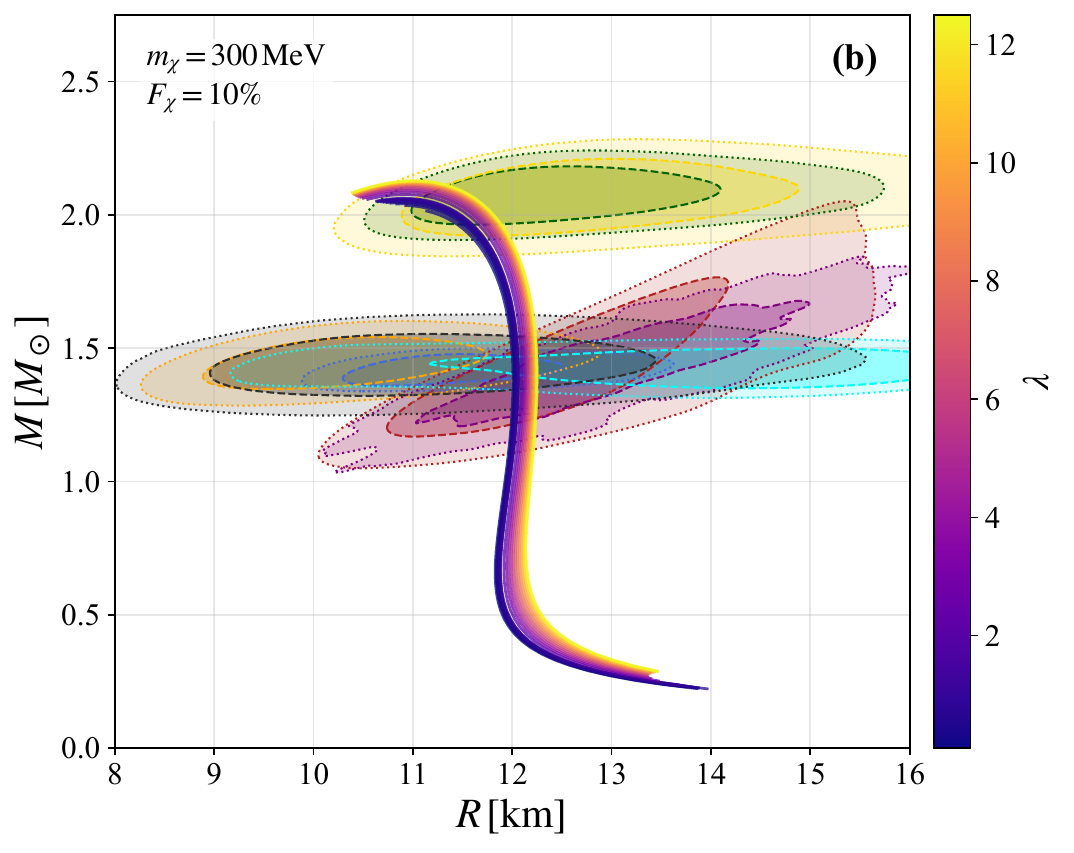}
        \label{fig:param_sensitivity_lambda}
    \hfill
        \includegraphics[width=0.32\linewidth]{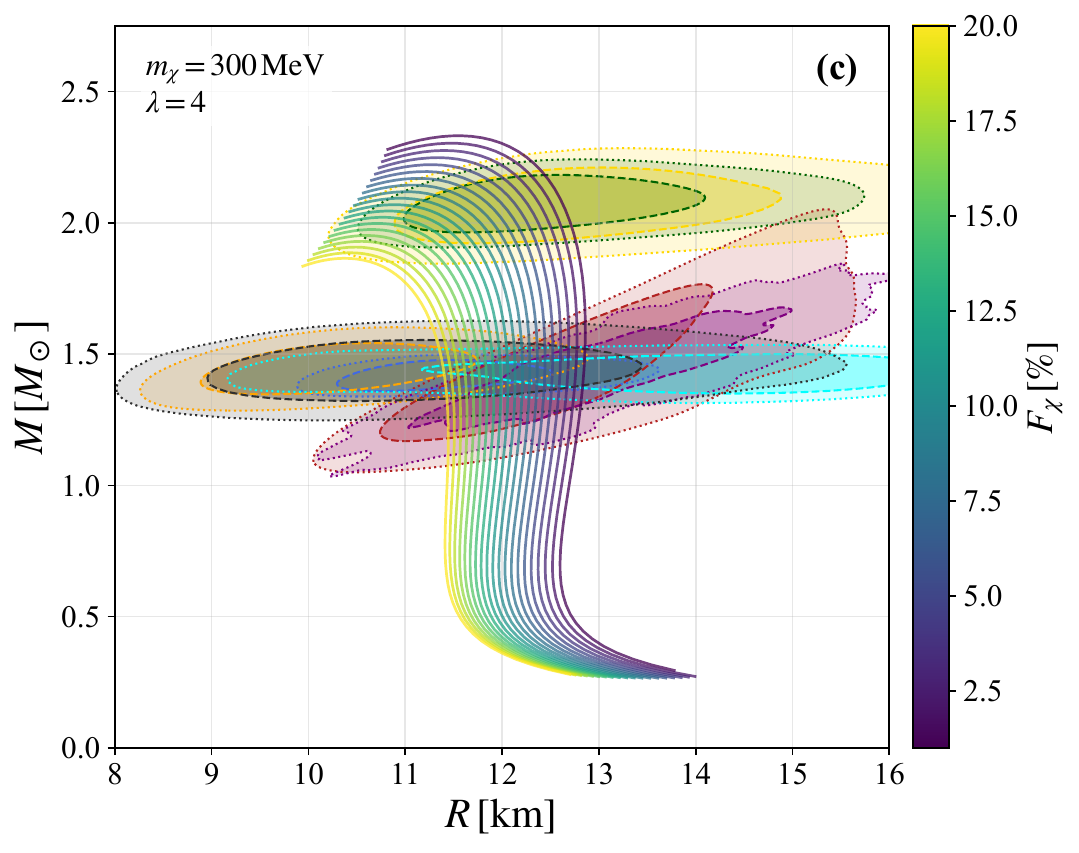}
        \label{fig:param_sensitivity_fchi}
    \caption{
    Sensitivity of the mass--radius relation to the individual bosonic-DM parameters
    for the representative baryonic EOS with $L_{\rm sym}=50~{\rm MeV}$ and
    $Q_{\rm sat}=0~{\rm MeV}$. In each panel, only one dark-sector parameter is
    varied while the remaining two are fixed: (a) varying $m_\chi$ at fixed
    $\lambda=4$ and $F_\chi=10\%$, (b) varying $\lambda$ at fixed
    $m_\chi=300~{\rm MeV}$ and $F_\chi=10\%$, and (c) varying $F_\chi$ at fixed
    $m_\chi=300~{\rm MeV}$ and $\lambda=4$. The colored curves show the resulting
    DM-admixed stellar sequences, with the colorbar indicating the varied
    parameter. The filled contours show the NICER mass--radius credible regions from both
    the Amsterdam and Maryland/Illinois analyses, while dashed and
    dotted boundaries denote the 68\% and 95\% credible regions, respectively.
    Within the particular one-parameter slices and parameter ranges shown here, variations in $F_\chi$ and $m_\chi$ produce larger   changes in the stellar sequences than the displayed variation in $\lambda$.
    The plots are shown in terms of the BM radius, $R\equiv R_{\rm BM}$.  }
    \label{fig:dm_parameter_sensitivity_mr}
\end{figure*}

Figure~\ref{fig:dm_parameter_sensitivity_mr} illustrates the sensitivity of the mass--radius relation to the three bosonic-DM parameters for a representative baryonic EOS. The purpose of this figure is not to define an excluded or allowed region, but rather to show how strongly each dark-sector parameter can deform the stellar sequence. Varying the DM fraction produces the largest change in the mass--radius plane, because $F_\chi$ directly controls how much of the total gravitational mass is carried by the dark component. As $F_\chi$ is changed, both the radius and the maximum supported mass are significantly affected.  The dependence on the boson mass is also substantial. Changing $m_\chi$ modifies the compactness and spatial distribution of the dark component, which in turn changes the mass--radius sequence. At fixed $m_\chi$ and $F_\chi$, varying $\lambda$ over the range shown produces a comparatively smaller displacement of the stellar curves than the displayed variation of $m_\chi$. This comparison refers only to the particular one-parameter slices and ranges shown in Fig.~\ref{fig:dm_parameter_sensitivity_mr}. Since the dark-sector EOS depends on $\lambda/m_\chi^4$, it should not be interpreted as an independent measure of the relative observational sensitivity to $m_\chi$ and $\lambda$.

Increasing the boson mass makes the dark component more compact for fixed $\lambda$ and $F_\chi$, favoring a core-like DM distribution. This increases the gravitational pull on the ordinary matter and leads to a stronger reduction of both the BM radius and the maximum supported mass. The effect of the DM fraction is even more direct: for the relatively heavy boson mass considered here, increasing $F_\chi$ increases the amount of compact DM inside the star and therefore enhances the reduction of the mass--radius sequence. By contrast, increasing the self-interaction coupling $\lambda$ provides additional repulsive pressure in the dark sector. This partially counteracts the softening induced by the dark component and leads to a smaller reduction, or equivalently a relative increase, of the maximum mass compared with weakly self-interacting configurations.
Since the dark-sector EOS depends on the combination $\lambda/m_\chi^4$, the individual variations of $m_\chi$ and $\lambda$ shown in Fig.~\ref{fig:dm_parameter_sensitivity_mr} correspond to different one-dimensional trajectories through the same effective EOS parameter space. Their relative impact therefore also reflects the different ranges sampled along these trajectories.

\begin{figure*}[t] \centering \includegraphics[width=0.95\textwidth]{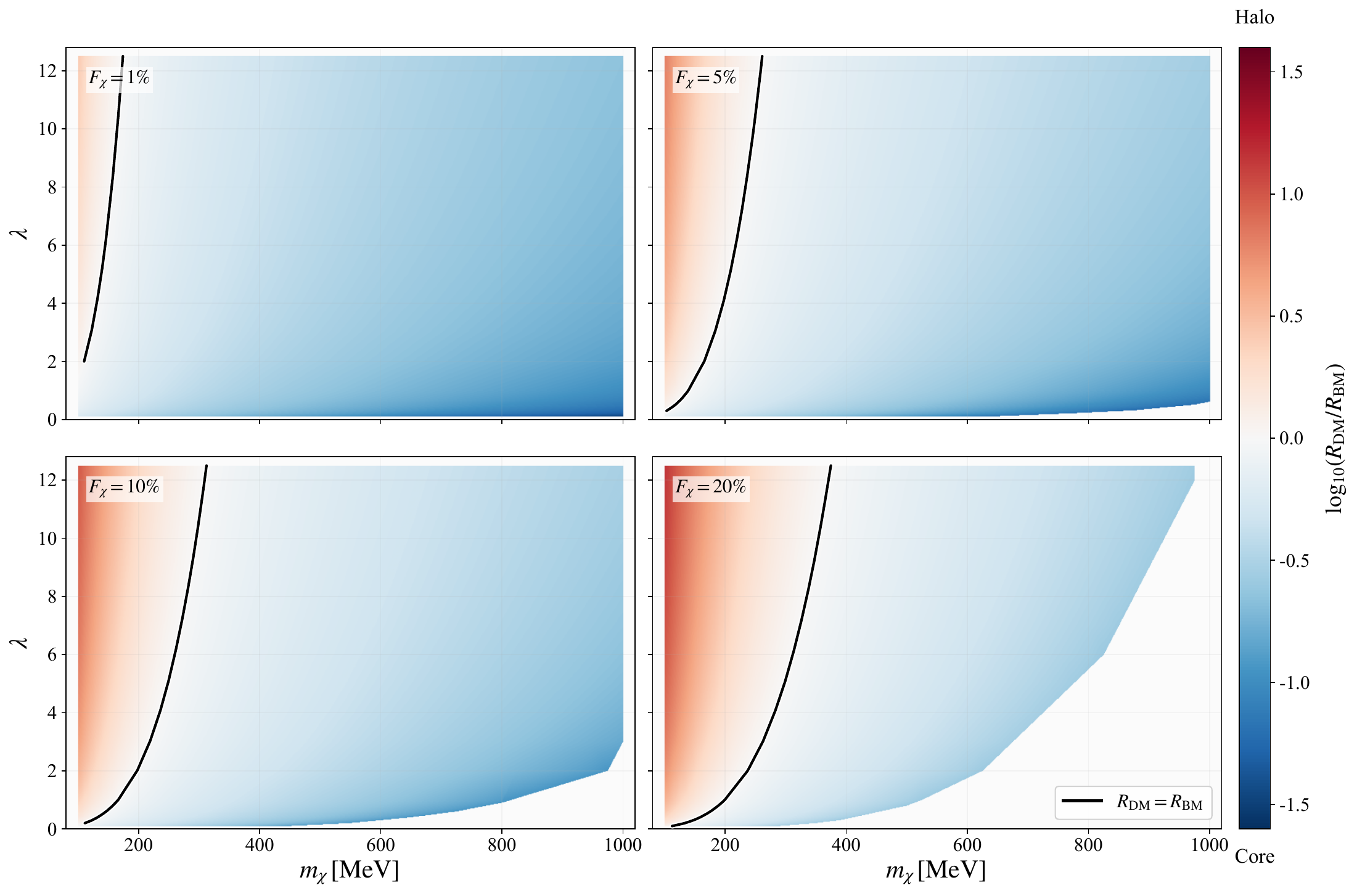} \caption{ {Core--halo structure of the bosonic-DM admixed $1.4M_\odot$ configurations for the representative EOS with $L_{\rm sym}=50$ MeV and $Q_{\rm sat}=0$ MeV. Each panel corresponds to a fixed dark fraction, $F_\chi=1\%,5\%,10\%$, or $20\%$. The color scale shows $\log_{10}(R_{\rm DM}/R_{\rm BM})$, with negative values corresponding to core-like configurations, $R_{\rm DM}<R_{\rm BM}$, and positive values to halo-like configurations, $R_{\rm DM}>R_{\rm BM}$. The black curve marks the transition $R_{\rm DM}=R_{\rm BM}$, corresponding to $\log_{10}(R_{\rm DM}/R_{\rm BM})=0$. White regions denote parameter values for which no $1.4M_\odot$ configuration is obtained.}}

\label{fig:core_halo_ratio_map}
\end{figure*}

\section{DM Core--Halo Distribution}\label{sec:DM_Core-Halo-distribution}

In this section, we examine how the bosonic-DM parameters determine the spatial distribution of the dark component and modify the global properties of the star. We first discuss the transition between dark-core and dark-halo configurations and then consider the corresponding impact on the maximum supported mass.

\subsection{Radius of DM-admixed NS}

Figure~\ref{fig:core_halo_ratio_map} summarizes the core--halo structure of the bosonic DM admixed NSs with $M=1.4M_\odot$ masses for the representative EOS with $L_{\rm sym}=50$ MeV and $Q_{\rm sat}=0$ MeV. The color scale indicates $\log_{10}(R_{\rm DM}/R_{\rm BM})$, with negative values corresponding to core-like configurations, $R_{\rm DM}<R_{\rm BM}$, and positive values corresponding to halo-like configurations, $R_{\rm DM}>R_{\rm BM}$. The black curve shows the DM core-halo transition $R_{\rm DM}=R_{\rm BM}$, for which $\log_{10}(R_{\rm DM}/R_{\rm BM})=0$. White regions correspond to parameter values for which no valid $1.4M_\odot$ object is obtained.

The halo-forming region is mainly associated with light bosons and sufficiently large self-couplings. This behavior can be understood from the dark-sector EOS. For smaller $m_\chi$, the dark fluid is effectively stiffer, or equivalently has larger pressure support, compared with the case of heavier bosons. A larger self-coupling further increases the pressure support in the dark sector. These two effects allow the dark fluid to extend beyond the baryonic radius and form a halo. At the smallest dark fraction, $F_\chi=1\%$, halo formation is restricted to very light bosons and relatively large self-couplings, and the resulting halos remain comparatively modest. As $F_\chi$ increases, the halo-forming region expands systematically: the transition boundary moves toward larger $m_\chi$, and halos appear over a wider range of $\lambda$. Thus, the most extended halos occur for light bosons, strong self-interactions, and large dark fractions. 

The core-like region shows the complementary behavior. Core formation is favored by heavier bosons and smaller self-couplings. Heavier bosons correspond to a softer dark-sector EOS, so the dark component is more easily confined inside the baryonic radius. At fixed $F_\chi$ and fixed $\lambda$, increasing $m_\chi$ therefore makes the dark component more compact and decreases $R_{\rm DM}/R_{\rm BM}$. At fixed $F_\chi$ and fixed $m_\chi$, increasing $\lambda$ works against strong core formation because the self-interaction provides additional pressure support in the dark sector.  Larger $\lambda$ therefore makes the dark distribution less compact and pushes $R_{\rm DM}/R_{\rm BM}$ closer to unity or equivalently $\log_{10}(R_{\rm DM}/R_{\rm BM})$ toward zero. The strongest core-like behavior is consequently obtained for large $m_\chi$ and small $\lambda$, with the effect becoming more pronounced as the dark fraction is increased.

Although the defining condition for a core-like configuration is
$R_{\rm DM}<R_{\rm BM}$, the physical effect of the core branch is not limited
to the fact that the dark component is confined inside the baryonic star. A
compact dark core also makes the visible baryonic component more compact. In
the core branch, the baryonic radius is the outermost radius of the
DM-admixed star, so a reduction of $R_{\rm BM}$ corresponds directly to a
more compact visible stellar configuration. For the same EOS without DM, the
radius of $M=1.4M_\odot$ star is
$R_{\rm BM}^{\rm noDM}=12.931~{\rm km}$. In the core-like region, the
baryonic radius is reduced relative to this no-DM value. This reduction is
mild at small dark fraction, about $0.08$--$0.10~{\rm km}$ at
$F_\chi=1\%$, but becomes much stronger at large dark fraction, reaching
about $1.5$--$2.0~{\rm km}$ for the strongest core-like configurations at
$F_\chi=20\%$. Thus, the core branch should be understood not only as a small
DM radius, but also as a more compact DM-admixed NS.

Figure~\ref{fig:core_halo_ratio_map} shows that lighter bosons and stronger self-interactions favor extended dark halos, whereas heavier bosons and weaker self-interactions favor compact dark cores. Increasing $F_\chi$ enhances these trends: in the halo branch it increases the radial extension of the dark component, while in the core branch it strengthens the compactifying effect and further reduces the baryonic radius. At fixed $1.4M_\odot$, this stronger core effect is also reflected by the growing white regions, where the stellar sequences no longer support a $1.4M_\odot$ configuration.
%------------------------------
\begin{figure*}[t]
\centering
\includegraphics[width=0.98\textwidth]{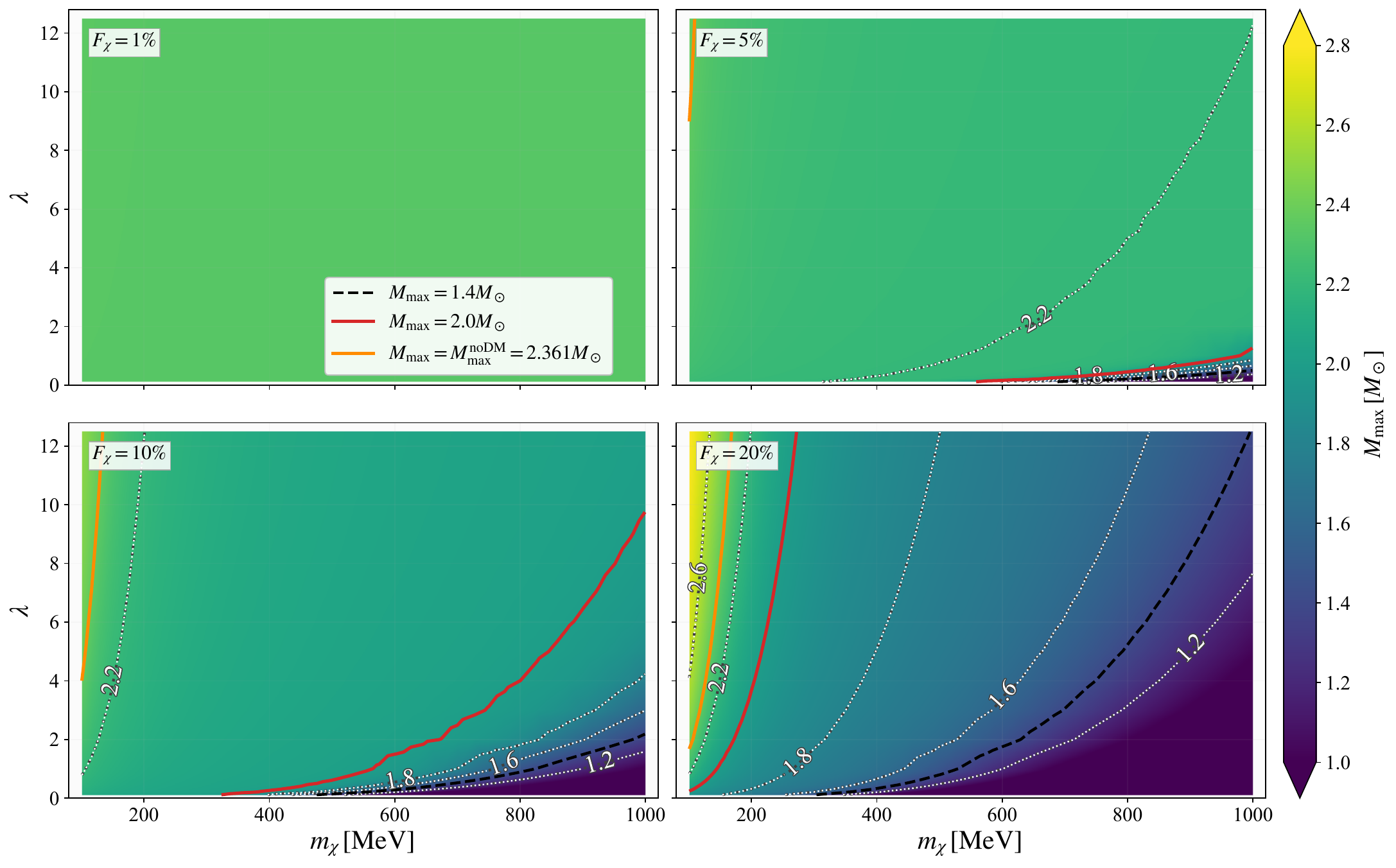}
\caption{
Maximum gravitational mass, $M_{\rm max}$, attained along the
bosonic-DM-admixed stellar sequences in the $(m_\chi,\lambda)$ plane
for the representative EOS with $L_{\rm sym}=50$ MeV and
$Q_{\rm sat}=0$ MeV. Each panel corresponds to a fixed DM
fraction, $F_\chi=1\%,\,5\%,\,10\%$, or $20\%$. The color scale shows
$M_{\rm max}$ in units of $M_\odot$; the lower and upper extensions
of the colorbar indicate values outside the plotted range,
$M_{\rm max}<1.0\,M_\odot$ and $M_{\rm max}>2.8\,M_\odot$,
respectively. The black dashed contour marks
$M_{\rm max}=1.4\,M_\odot$, below which the sequence does not contain
a $1.4\,M_\odot$ configuration. The red solid contour marks
$M_{\rm max}=2.0\,M_\odot$, corresponding to the conventional
two-solar-mass requirement. The orange contour marks
$M_{\rm max}=M_{\rm max}^{\rm noDM}=2.361\,M_\odot$ and therefore
defines the maximum-mass crossover between regions in which the dark
component decreases or increases the maximum mass relative to the
corresponding purely baryonic sequence. Additional dotted contours
indicate intermediate values of $M_{\rm max}$.
}
\label{fig:mmax_parameter_map}
\end{figure*}
%---------------------------------------

\subsection{Maximum mass of DM-admixed NS}

Figure~\ref{fig:mmax_parameter_map}  shows the maximum gravitational mass attained along the constructed fixed-$F_\chi$ equilibrium sequences. The condition $M_{\max}^{\rm DM}=M_{\max}^{\rm noDM}$ defines a maximum-mass crossover: on its two sides, the dark component decreases or increases the maximum mass attained along the sequence relative to the corresponding no-DM model. The standard $2M_\odot$ requirement is shown by the red contour, while the black dashed $M_{\max}=1.4M_\odot$ contour is included as an auxiliary guide to connect this plot with the previous fixed-$1.4M_\odot$ core--halo map where the white regions in Fig. \ref{fig:core_halo_ratio_map} correspond to the sequences that do not contain a $1.4M_\odot$ configuration.

For the same EOS without DM, the maximum mass is
$M_{\max}^{\rm noDM}=2.361M_\odot$. At the smallest dark fraction,
$F_\chi=1\%$, the effect of the dark component on the maximum mass is weak,
and the sequences remain close to the no-DM value. As the dark
fraction increases, however, the dependence of $M_{\max}$ on $m_\chi$ and
$\lambda$ becomes much stronger.

The largest maximum masses occur in the region of light bosons and large self-couplings. This behavior is consistent with the dark-sector EOS interpretation discussed above: lighter bosons and stronger self-interactions provide greater pressure support in the dark component, allowing the admixed star to support larger masses. Relative to the no-DM value, the increase is small at $F_\chi=5\%$, but it becomes more pronounced at larger dark fractions, as shown for $F_\chi=10\%$ and 20\%. The highest maximum masses occur near the lightest boson masses and largest self-couplings, where the dark component provides the greatest pressure support.

The opposite behavior appears for heavy bosons and small self-couplings. In this region, the dark-sector EOS is softer, the dark component becomes more compact, and the compact dark core reduces the mass that can be supported by the stellar sequence. This effect becomes progressively stronger with increasing $F_\chi$. The lowest maximum mass values occur in the heavy-boson, weakly self-interacting region of the scan and are far below both the no-DM maximum mass and the $2M_\odot$ requirement.

Figure~\ref{fig:mmax_parameter_map} confirms that the maximum-mass behavior is governed by the effective combination $\lambda/m_\chi^4$. Larger values of $\lambda/m_\chi^4$, corresponding to lighter bosons and/or stronger repulsive self-interactions, provide greater pressure support in the dark sector and can yield maximum masses comparable to, or even larger than, the no-DM value. Smaller values of $\lambda/m_\chi^4$, corresponding to heavier bosons and/or weaker self-interactions, favor more compact dark cores and can strongly reduce the maximum mass attained along the stellar sequence. The dark fraction $F_\chi$ amplifies both effects.

It is important to distinguish the core--halo transition from the transition in the maximum-mass behavior. The condition $R_{\rm DM}=R_{\rm BM}$ is a structural criterion: it marks the point at which the dark component changes from being confined inside the baryonic radius to forming an extended halo. By contrast, the condition $M_{\max}^{\rm DM}=M_{\max}^{\rm noDM}$ defines the maximum-mass crossover, at which the maximum mass attained along the DM-admixed sequence equals that of the corresponding purely baryonic sequence.  These two boundaries are therefore not expected to coincide. A halo can form once the dark component has enough pressure support to extend beyond the baryonic surface, but at this stage the halo may still be too dilute, or the reduction of the baryonic contribution may still be too strong, for the total maximum mass to exceed the no-DM value. Only after the dark component provides enough additional pressure support does it compensate for the baryonic compactification and lead to $M_{\max}^{\rm DM}>M_{\max}^{\rm noDM}$.  This behavior is consistent with the core--halo transition found in Ref.~\cite{Karkevandi:2021ygv}, where the onset of $R_{\rm DM}>R_{\rm BM}$ occurred before the maximum mass began to increase.  Thus, the present core--halo and maximum-mass maps show two related but physically distinct thresholds: the first describes the spatial distribution of the dark component, while the second describes its net effect on the maximum mass attained along the sequence.

\section{Bayesian Analysis}\label{sec:Bayesian_analysis}

We perform a Bayesian analysis to constrain the three dark-sector parameters of the bosonic-DM model and to quantify their degeneracy with the baryonic EOS. The analysis is carried out separately for each of the 27 nucleonic EOSs introduced above and is subsequently combined through Bayesian model averaging.

\subsection{Parameter Space and Prior Distributions}

Let $E_i$, with $i=1,\ldots,27$, denote a member of the baryonic EOS ensemble, characterized by a fixed pair $(L_{\rm sym},Q_{\rm sat})$. The dark-sector parameters are collected in the vector $\theta=(m_\chi,\lambda,F_\chi)$, where $m_\chi$ is the boson mass, $\lambda$ is the self-interaction coupling constant, and $F_\chi$ is the DM fraction of the mixed object.

For every baryonic EOS $E_i$, we evaluate the stellar models on the same three-dimensional parameter grid $\Theta=\{\theta_{abc}=(m_\chi^{(a)},\lambda^{(b)},F_\chi^{(c)})\}$,
where $a=1,\ldots,55$, $b=1,\ldots,22$, $c=1,\ldots,20$.
The grid therefore contains $N_\theta=55\times22\times20=24\,200$ dark-sector configurations for each fixed baryonic EOS.

We assign independent discrete uniform priors to the sampled values of $m_\chi$, $\lambda$, and $F_\chi$. Consequently, every sampled triplet carries the same prior probability,
$\pi(\theta_{abc})=1/N_\theta=1/{24\,200}$.

\subsection{Multimessenger Constraints and Likelihood Construction}

For a given NICER data set $X$, the complete observational data set is denoted by 
\begin{equation}
D_X = \left\{ D_{\rm GW}, \left\{D_X^{(s)}\right\}_{s\in\mathcal S_X} \right\},
\label{eq:complete_data_set}
\end{equation}
where  $D_{\rm GW}$ denotes the GW170817 constraint, and $D_X^{(s)}$ denotes the NICER data for source $s$. The label $X$ distinguishes between the Amsterdam and Maryland/Illinois NICER analyses, while $\mathcal S_X$ denotes the corresponding set of NICER sources.

Assuming that the individual measurements are statistically independent, the total likelihood factorizes as 
\begin{align}
\mathcal{L}(D_X|\theta,E_i)
={}&
\mathcal{L}_{\rm GW}(D_{\rm GW}|\theta,E_i)
\nonumber\\[2mm]
&\times
\prod_{s\in\mathcal{S}_X} \mathcal{L}_{\rm NICER} \left(D_X^{(s)}|\theta,E_i\right).
\label{eq:total_likelihood}
\end{align}

\subsubsection{NICER Mass--Radius Constraints}

In order to examine the dependence of the inferred dark-sector parameters on the adopted X-ray pulse-profile analysis, we consider the Amsterdam and Maryland/Illinois NICER results separately. For the Amsterdam analysis, the source set contains PSR~J0030+0451~\cite{Kini:2026rjx}, PSR~J0740+6620~\cite{Salmi:2024aum}, PSR~J0437$-$4715~\cite{Choudhury:2024xbk}, and PSR~J0614$-$3329~\cite{Mauviard:2025dmd}. For the Maryland/Illinois analysis, the source set contains PSR~J0030+0451~\cite{Miller:2019cac}, PSR~J0740+6620~\cite{Dittmann:2024mbo}, PSR~J0437$-$4715~\cite{Miller:2025qfq} and PSR~J0614$-$3329~\cite{Miller:2026vpr} .

For every NICER source $s$, we construct a continuous approximation to the published two-dimensional mass--radius posterior by applying a kernel-density estimator (KDE) to the publicly available posterior samples. For the Amsterdam analysis, we use the samples for 
PSR~J0030+0451~\cite{kini_2026_20271881},
PSR~J0740+6620~\cite{salmi_2024_10519473},
PSR~J0437--4715~\cite{choudhury_2024_13766753}, and
PSR~J0614--3329~\cite{mauviard_2025_17380576}. 
For the Maryland/Illinois analysis, we use the corresponding samples for
PSR~J0030+0451~\cite{miller:2019:3473466},
PSR~J0740+6620~\cite{dittmann_2024_10215109},
PSR~J0437--4715~\cite{miller_2026_17833896} and PSR~J0614--3329~\cite{miller_2026_22131748}.

For defining the line integral, we introduce the dimensionless coordinates $\widetilde M=M/M_\odot$ and $\widetilde R=R/1\,{\rm km}$. For a model specified by $(\theta,E_i)$, the continuous equilibrium mass–radius sequence is 
\begin{equation}
\widetilde{\mathcal C}_{i,\theta}^{\,MR} =\left\{\left(\widetilde M,\widetilde R_{i,\theta}(\widetilde M)\right)\;\middle|\;\widetilde M\in\left[0.2,\widetilde M_{\max}(\theta,E_i)\right]\right\}.
\label{eq:dimensionless_mr_curve}
\end{equation}
The NICER likelihood for source $s$ is calculated as the curvilinear integral of the KDE-reconstructed posterior density along this sequence:
\begin{equation}
\mathcal{L}_{\rm NICER}
\left(D_X^{(s)}|\theta,E_i\right) = \int_{\widetilde{\mathcal C}_{i,\theta}^{\,MR}} \mathrm{KDE}^{(s)}_X \left(\widetilde M,\widetilde R\right)
\,d\widetilde{\ell}_{MR},
\label{eq:nicer_line_integral}
\end{equation}
where the dimensionless arclength element is $d\widetilde{\ell}_{MR}=\sqrt{d\widetilde M^{\,2}+d\widetilde R^{\,2}}$.

\subsubsection{GW170817 Tidal-Deformability Constraint}

The tidal-deformability constraint is constructed analogously to the NICER mass–radius constraints. We use the publicly available posterior samples for GW170817 released by the LIGO-Virgo Collaborations through the LIGO Document Control Center~\cite{ligoLIGOP1800115v12GW170817}. A two-dimensional KDE is applied to the $(\Lambda_1,\Lambda_2)$ samples to obtain the continuous density function $\mathrm{KDE}_{\rm GW}(\Lambda_1,\Lambda_2)$.

The accurately measured source-frame chirp mass of GW170817 is $\mathcal M_{\rm chirp} = 1.188^{+0.004}_{-0.002}\,M_\odot$ 
at the 90\% credible level~\cite{Abbott_2017}. Since its uncertainty is small compared with those of the individual component masses, we fix $\mathcal M_{\rm chirp}=1.188\,M_\odot$ when constructing the theoretical $\Lambda_1$--$\Lambda_2$ curve. 
For a binary mass ratio $q=M_2/M_1\leq1$, the component masses are
\begin{align}
M_1(q) &= \mathcal M_{\rm chirp} (1+q)^{1/5}q^{-3/5}, \nonumber\\
M_2(q) &= \mathcal M_{\rm chirp} (1+q)^{1/5}q^{2/5}.
\label{eq:gw_component_masses}
\end{align}

The observed low-spin interval is taken to be $q_{\min}^{\rm obs}=0.7$ and $q_{\max}^{\rm obs}=1$ 
following Ref.~\cite{Abbott_2017}. The mass-ratio interval used for the model is therefore $\mathcal I_q^{\,i,\theta} = \left[q_{\min}(\theta,E_i),1\right]$, where $q_{\min}(\theta,E_i)=\max\left[q_{\min}^{\rm obs}, q_{\min}^{\rm EOS}(\theta,E_i)\right]$.

For the fixed chirp mass, the minimum possible primary mass occurs for an equal-mass binary, $q=1$, and is
$M_{1,\min}=2^{1/5}\mathcal M_{\rm chirp}\simeq 1.36\,M_\odot$.
Consequently, the GW170817 likelihood is set to zero when
$M_{\max}(\theta,E_i)<1.36\,M_\odot$, because such a model cannot support the primary component of GW170817 for any $q\leq1$.

For models satisfying the  condition, the theoretical
tidal-deformability curve is defined as
\begin{equation}
\mathcal C_{i,\theta}^{\,\Lambda} = \left\{\left({\Lambda_1}_{i,\theta}[M_1(q)],{\Lambda_2}_{i,\theta}[M_2(q)]\right)\;\middle|\;q\in\mathcal I_q^{\,i,\theta}\right\}.
\label{eq:lambda_curve}
\end{equation}
The GW170817 likelihood is calculated as the curvilinear integral of the KDE-reconstructed density along this curve:
\begin{equation}
\mathcal{L}_{\rm GW}(D_{\rm GW}|\theta,E_i) = 
\int_{\mathcal C_{i,\theta}^{\,\Lambda}}\mathrm{KDE}_{\rm GW}\left(\Lambda_1,\Lambda_2\right)\,d\ell_\Lambda,
\label{eq:gw_line_integral}
\end{equation}
where $d\ell_\Lambda = \sqrt{d\Lambda_1^{\,2} + d\Lambda_2^{\,2}}$. We use the KDE reconstructions of the published observational posteriors as effective likelihood densities and apply this prescription uniformly to all EOS and dark-sector models. 

Following the curvilinear-overlap approach employed in previous Bayesian analyses of NS EOSs
\cite{Brandes:2023,Ayriyan:2026}, we evaluate the observational constraints by integrating KDE representations of the published posteriors along the corresponding theoretical sequences.
We emphasize that the curvilinear construction adopted here represents a specific choice of measure along the theoretical sequence. Since the integration is performed with the arclength element, it is invariant under a reparameterization of a fixed curve, but it depends on the metric chosen in the corresponding observable space. In particular, for the mass--radius likelihood this metric is defined by the dimensionless coordinates $\widetilde M=M/M_\odot$ and $\widetilde R=R/(1\,{\rm km})$. Consequently, the resulting effective likelihood reflects not only whether the theoretical sequence passes through a region of high posterior density, but also the extent of the sequence within that region. We regard this geometrical weighting as
part of the adopted effective-likelihood prescription and apply it consistently to all EOS and dark-sector models.

\subsection{Inference and Evidence}

Having specified the prior distributions and the complete multimessenger likelihood, we first perform the Bayesian inference separately for each baryonic EOS $E_i$. At a sampled grid point $\theta_{abc}$, the fixed-EOS posterior probability is
\begin{equation}
p(\theta_{abc}|D_X,E_i) = \frac{\mathcal L(D_X|\theta_{abc},E_i)\,\pi(\theta_{abc})}{Z_i^X},
\label{eq:fixed_eos_posterior}
\end{equation}
where the evidence is evaluated as the discrete sum $Z_i^X = \sum_{a}\sum_{b}\sum_{c} \mathcal L(D_X|\theta_{abc},E_i)\,\pi(\theta_{abc})$.

The posterior probability assigned to baryonic EOS $E_i$ is given as 
\begin{equation}
p(E_i|D_X) = \frac{Z_i^X\pi(E_i)}{\displaystyle\sum_{j=1}^{27}Z_j^X\pi(E_j)}.
\label{eq:eos_posterior}
\end{equation}
Since all EOSs are assigned equal prior probabilities, $\pi(E_i)=1/27$, the normalized EOS weights reduce to
\begin{equation}
w_i^X = p(E_i|D_X) = Z_i^X \Big/ \displaystyle \sum_{j=1}^{27}Z_j^X.
\label{eq:eos_weights}
\end{equation}

The EOS-marginalized posterior distribution of the dark sector parameters is then
\begin{equation}
p(\theta_{abc}|D_X) = \sum_{i=1}^{27} p(\theta_{abc}|D_X,E_i)\, w_i^X.
\label{eq:model_averaged_posterior}
\end{equation}
Thus, EOSs that are more compatible with the observational data contribute more strongly to the final posterior. The same evidence-based weights are used to infer the posterior probabilities of the nuclear parameters $(L_{\rm sym},Q_{\rm sat})$ and to compare the no-DM and DM-admixed model families. This procedure incorporates the uncertainty associated with the discrete baryonic EOS ensemble into the inferred constraints on the bosonic-DM parameters.

\subsection{Posterior-weighted mass--radius sequences}

\begin{figure*}[t]
    \centering
    \includegraphics[width=0.48\textwidth]{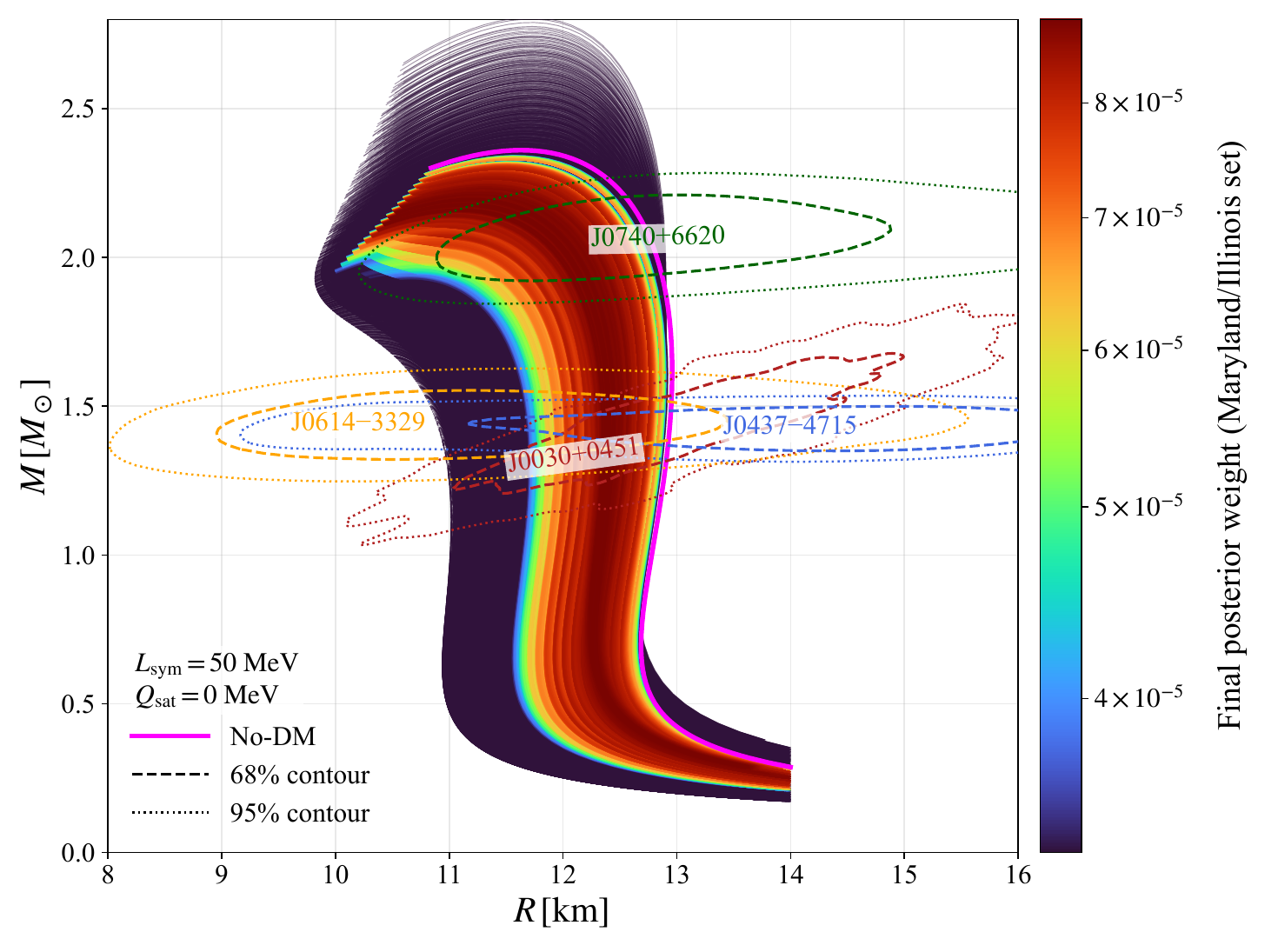}
    \hfill
    \includegraphics[width=0.48\textwidth]{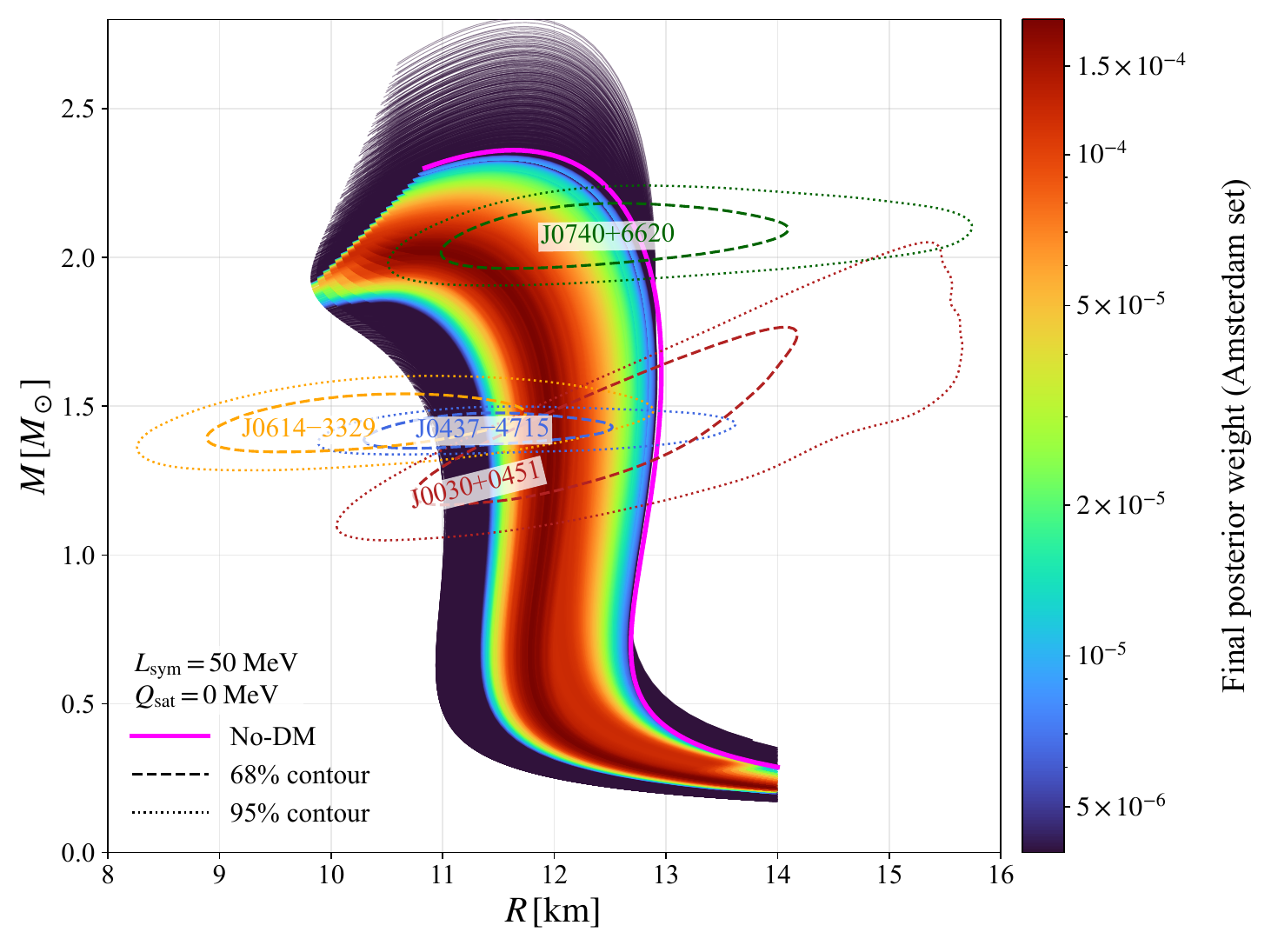}
    \caption{
    Posterior-weighted mass--radius relations of bosonic-DM admixed NSs for the representative baryonic EOS with \(L_{\rm sym}=50~{\rm MeV}\) and \(Q_{\rm sat}=0~{\rm MeV}\).  Each colored curve corresponds to one DM-admixed stellar sequence in the scanned dark-sector parameter space for this fixed EOS, with the color indicating the final normalized posterior weight for this specific EOS obtained after applying the corresponding NICER mass--radius likelihood together with the GW170817 tidal-deformability constraint.   The magenta curve shows the no-DM sequence. 
Dashed and dotted contours denote the 68\% and 95\% NICER Bayesian posterior regions, respectively.     Left panel: Maryland/Illinois NICER analysis.  Right panel: Amsterdam NICER analysis.  The color scale is percentile-clipped for visualization. The plots are shown in terms of the BM
radius, $R\equiv R_{\rm BM}$.
}
 \label{fig:MR_posterior_weighted}
\end{figure*}

To illustrate how the Bayesian weighting translates into the
mass--radius plane, Fig.~\ref{fig:MR_posterior_weighted}  shows the posterior-weighted stellar
sequences for a representative baryonic EOS with
$L_{\rm sym}=50~\mathrm{MeV}$ and $Q_{\rm sat}=0~\mathrm{MeV}$.

Each colored curve corresponds to one DM-admixed stellar sequence in the scanned dark-sector parameter space for this fixed EOS. The color indicates the final normalized posterior weight for this specific EOS, obtained from the Bayesian analysis described above after applying the corresponding NICER mass--radius likelihood together with the GW170817 tidal-deformability constraint. The magenta curve shows the corresponding no-DM sequence.

The comparison between the two panels illustrates the different preferences induced by the two NICER analyses. The Maryland/Illinois posterior assigns the largest weights to configurations that remain relatively close to the no-DM sequence, while the Amsterdam posterior gives more weight to configurations shifted farther away from the no-DM relation. This behavior anticipates the posterior constraints discussed below in Sec. \ref{sec:Results}, where the preferred boson mass remains in the few-hundred-MeV range for both analyses, whereas the preferred DM fraction is smaller for the Maryland/Illinois set and larger for the Amsterdam set.

An important feature of Fig.~\ref{fig:MR_posterior_weighted} is that the inclusion of the bosonic-DM component enlarges the set of mass--radius configurations accessible to a fixed baryonic EOS. For the representative no-DM sequence shown here, the purely baryonic curve does not pass through the most compact NICER regions, in particular those associated with the recent Amsterdam analysis. Once the dark component is included, however, part of the DM-admixed family is shifted toward smaller radii in the mass range relevant for these compact NICER constraints. Some of these shifted configurations retain sizable posterior weight, meaning that they are not selected by the mass--radius data alone, but also remain compatible with the GW170817 tidal-deformability constraint entering the full likelihood.

This behavior illustrates how bosonic DM can partially compensate for the limitations of a purely baryonic interpretation for a given EOS. In the no-DM case, the stellar sequence is fixed once $L_{\rm sym}$ and $Q_{\rm sat}$ are chosen. In contrast, allowing a bosonic-DM admixture introduces the additional parameters $\lambda,m_\chi,$ and $F_\chi$, which continuously deform the mass--radius relation. The posterior weighting then identifies those deformations that improve the simultaneous agreement with the compact NICER mass--radius regions and the tidal-deformability bound. This effect is especially visible for the Amsterdam data set, where the high-posterior DM-admixed curves are displaced farther from the no-DM sequence, consistent with the larger preferred DM fraction inferred from the posterior distribution. For the Maryland/Illinois data set, the preferred fraction is smaller, and the high-posterior curves remain closer to the purely baryonic sequence.

\section{Results}\label{sec:Results}

\subsection{Evidence-weighted constraints on the bosonic-DM parameter space}
\label{sec:parameter_space}

\begin{figure*} \centering \includegraphics[width=0.9\textwidth]{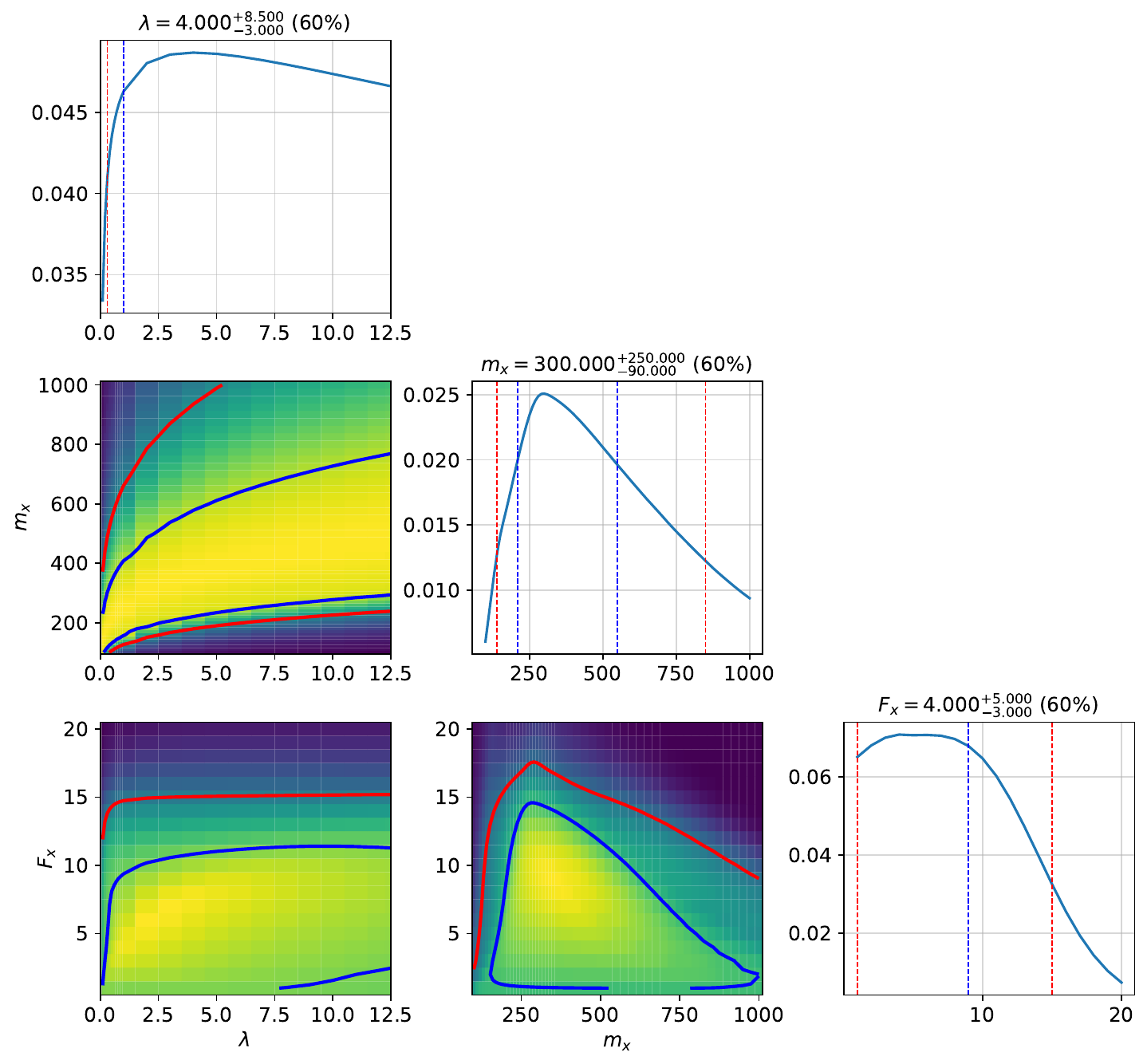} \caption{ Evidence-weighted model-averaged posterior distribution of the bosonic-DM parameters $(\lambda,m_\chi,F_\chi)$ obtained using the Maryland/Illinois NICER mass–radius analysis. The posterior combines the results of the 27 nucleonic EOSs through their Bayesian evidences. The diagonal panels show the one-dimensional marginalized posteriors, while the off-diagonal panels show the two-dimensional marginalized distributions. The blue and red contours denote the 60\% and 90\% highest-posterior-density regions, respectively. Regions outside the 90\% highest posterior density (HPD) contours correspond to the least probable regions, i.e., those outside the 90\% HPD contours.   }
\label{fig:post_DM_miller}
\end{figure*}

\begin{figure*}
\centering
\includegraphics[width=0.9\textwidth]{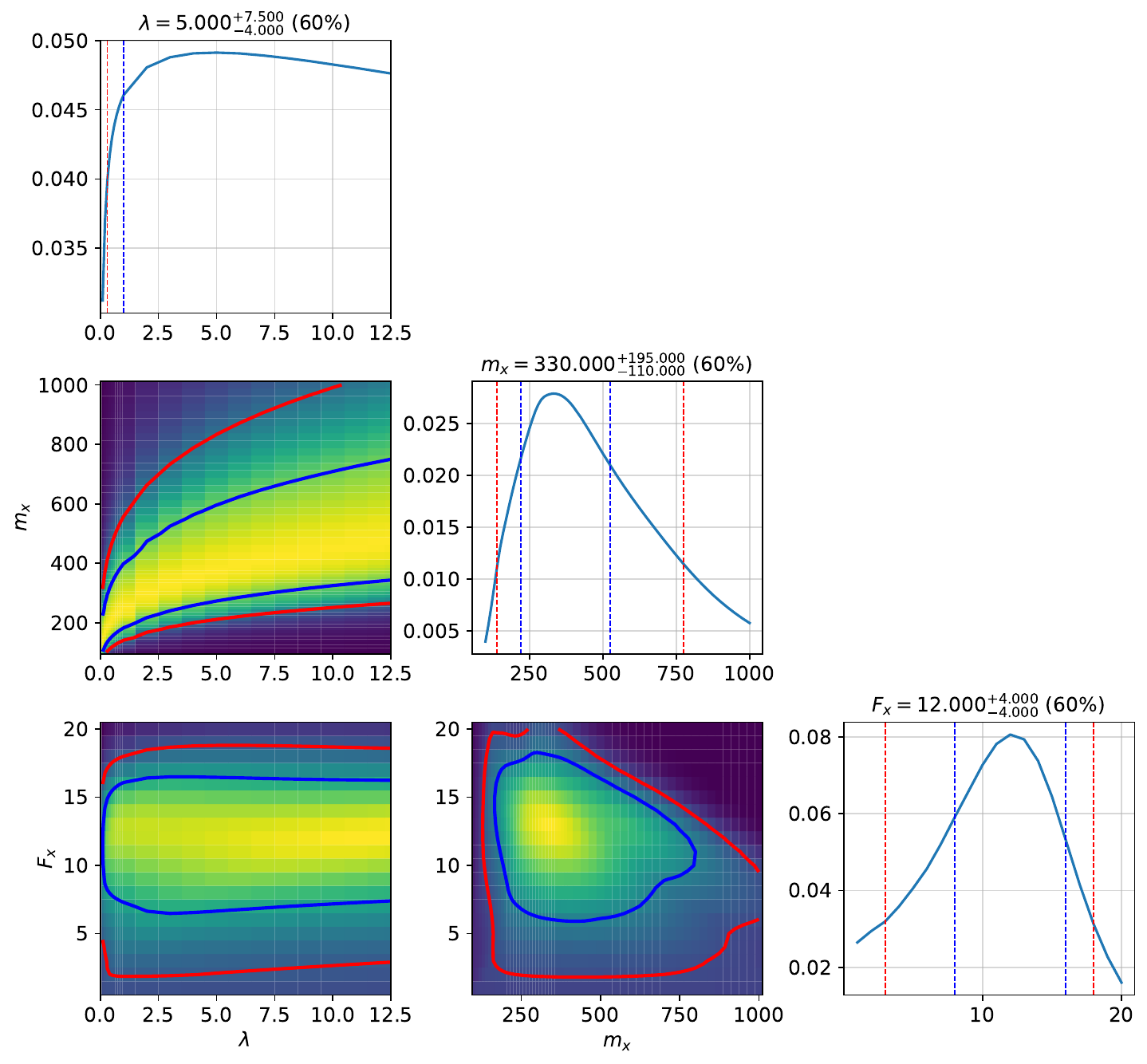}
\caption{
Same as Fig.~\ref{fig:post_DM_miller}, but using the Amsterdam NICER
mass–radius analysis. Compared with the Maryland/Illinois analysis, the
preferred boson mass remains in the few-hundred-MeV range, while the posterior
shifts toward a larger DM fraction. Regions outside the 90\% HPD contours
represent the least probable regions, i.e., those outside the 90\% HPD contours.
}
\label{fig:post_DM_watts}
\end{figure*}

\begin{figure*}[t]
\centering
\includegraphics[width=\textwidth]{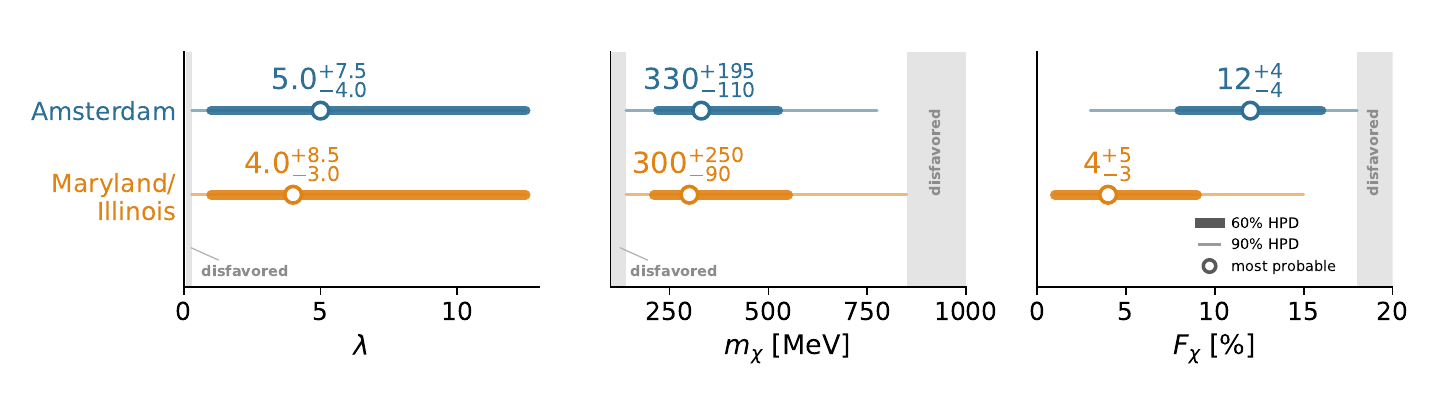}
\caption{%
Evidence-weighted marginalized constraints on the bosonic-DM parameters, the self-coupling $\lambda$, the boson mass $m_\chi$, and the DM fraction $F_\chi$, for the Maryland/Illinois (orange) and Amsterdam (blue) NICER mass–radius analyses. Open circles mark the most probable value, thick bars the 60\% HPD interval, and thin bars the 90\% HPD interval, all obtained from the one-dimensional marginalized posteriors of Figs.~\ref{fig:post_DM_miller} and~\ref{fig:post_DM_watts} after combining the 27 nucleonic EOSs through their Bayesian evidences. Grey shading marks the regions lying outside the 90\% HPD intervals, which are the least probable parts of the scanned parameter space. 
}
\label{fig:dm_hpd_forest}
\end{figure*}

Figures~\ref{fig:post_DM_miller} and~\ref{fig:post_DM_watts} show the evidence-weighted model-averaged posterior distributions for the bosonic-DM parameters $(\lambda,m_\chi,F_\chi)$. For each point in the dark-sector parameter space, the stellar configurations are computed for the 27 nucleonic EOSs and confronted with the observational constraints. The individual EOS-dependent results are then combined using their Bayesian evidences. Therefore, these posteriors do not correspond to a single assumed baryonic EOS, but represent the final constraints after accounting for the uncertainty within the selected CDF EOS ensemble. Figure~\ref{fig:dm_hpd_forest} summarizes the resulting one-dimensional marginalized constraints for both analyses side by side, and is referred to throughout the discussion below.

\begin{figure*}[t]
\centering
\includegraphics[width=0.96\textwidth]{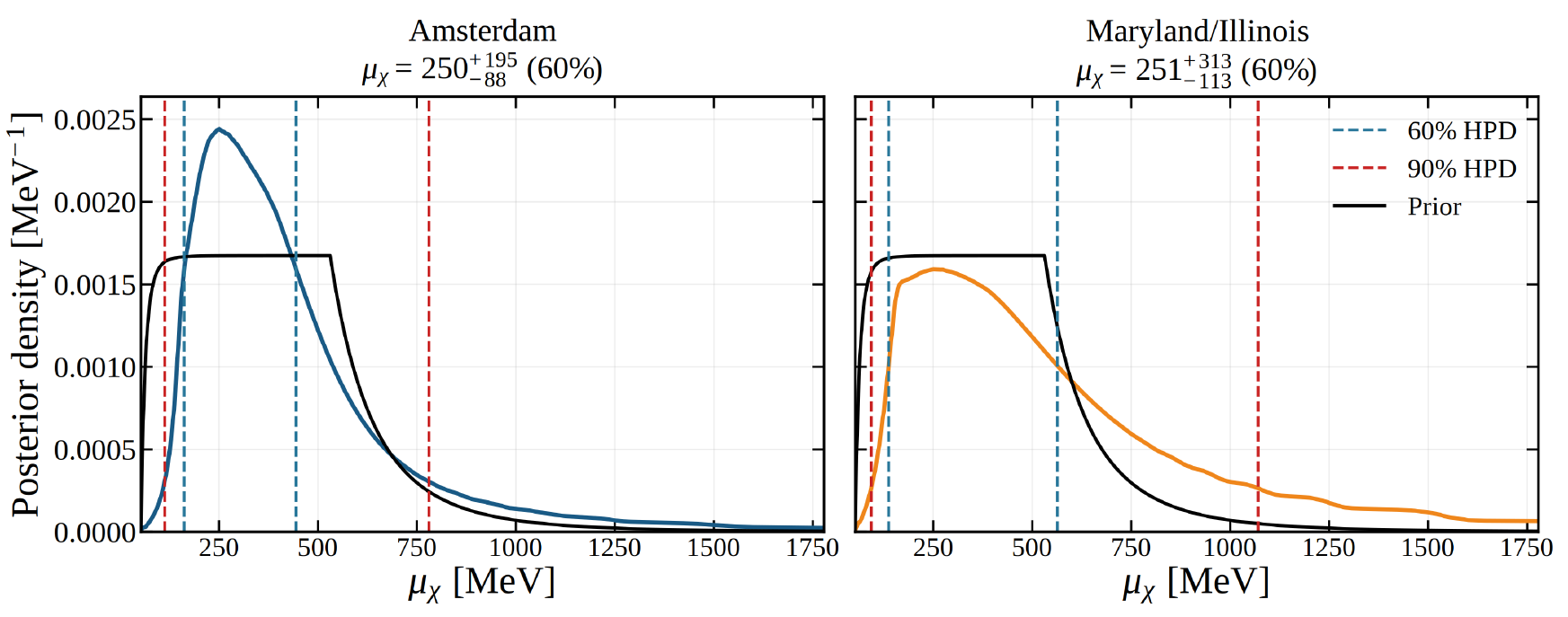}
\caption{%
Posterior probability density of the derived dark-sector scale $\mu_\chi=m_\chi/\lambda^{1/4}$ for the Amsterdam (left) and Maryland/Illinois (right) NICER analyses. The colored curves show the evidence-weighted posteriors after transforming the $(m_\chi,\lambda,F_\chi)$ posterior and marginalizing over $F_\chi$, while the black curve shows the prior on $\mu_\chi$ induced by the adopted independent discrete-uniform priors on $m_\chi$ and $\lambda$. Dashed vertical lines mark the 60\% and 90\% HPD intervals. The posterior modes and 60\% HPD intervals are $\mu_\chi=250^{+195}_{-88}$~MeV for Amsterdam and $\mu_\chi=251^{+313}_{-113}$~MeV for Maryland/Illinois. The Amsterdam posterior is more strongly localized relative to the induced prior, whereas the Maryland/Illinois posterior is broader and retains a longer tail toward large $\mu_\chi$.
}
\label{fig:mu_chi_distribution}
\end{figure*}

\begin{figure*}[t]
\centering
\includegraphics[width=0.6\textwidth]{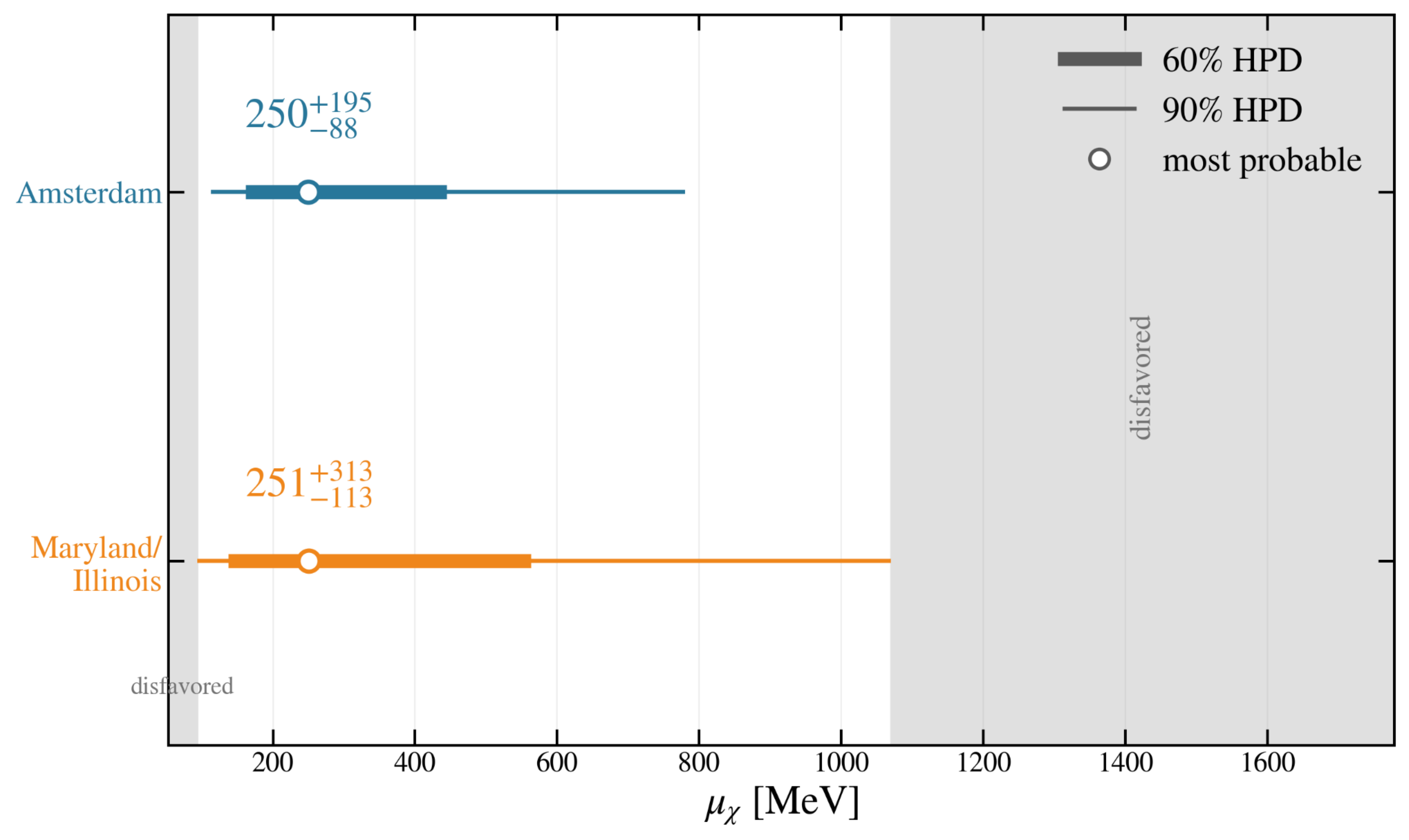}
\caption{%
Compact comparison of the inferred $\mu_\chi$ scale for the Amsterdam (blue) and Maryland/Illinois (orange) NICER analyses. Open circles mark the posterior modes, thick horizontal bars the 60\% HPD intervals, and thin horizontal bars the 90\% HPD intervals. The grey shading indicates regions lying outside the 90\% HPD intervals of both analyses. The two posteriors peak at essentially the same value, $\mu_\chi\simeq250$~MeV, while the Maryland/Illinois constraint is appreciably broader toward large $\mu_\chi$.
}
\label{fig:mu_chi_summary}
\end{figure*}

The diagonal panels show the one-dimensional marginalized posteriors for each parameter, while the off-diagonal panels display the corresponding two-dimensional marginalized distributions. These projections summarize the correlations among the dark-sector parameters after the full three-dimensional posterior has been marginalized over the remaining degrees of freedom. The color scale indicates the posterior probability density. The blue contours denote the 60\% highest-posterior-density (HPD) regions, which identify the most probable parts of the parameter space. The red contours denote the 90\% HPD regions, which give a broader credible region. Consequently, regions outside the 90\% HPD contours, or outside the 90\% intervals in the one-dimensional marginalized distributions, are the least probable regions of the scan and are referred to below as disfavored regions of the evidence-weighted posterior.  We note that, because of the $m_\chi–\lambda$ degeneracy in the dark EOS, these individual marginals should be read together with the derived $\mu_\chi$ posterior shown in Figs.~\ref{fig:mu_chi_distribution} and \ref{fig:mu_chi_summary}, which is less sensitive to specific parameterization. Note that we use 60\% and 90\% HPD regions as a reporting convention, with the former identifying the highest-posterior core and the latter a broader credible region. The conclusions are not tied to these particular credible levels.

For the Maryland/Illinois NICER analysis, shown in Fig.~\ref{fig:post_DM_miller}, the evidence-weighted posterior gives
\begin{equation}
\lambda=4.0^{+8.5}_{-3.0},\ 
m_\chi=300^{+250}_{-90}~{\rm MeV},\ 
F_\chi=4^{+5}_{-3}\%,
\label{eq:maryland_60}
\end{equation}
at the 60\% HPD level. This means that the most probable region is centered around a boson mass of a few hundred MeV and a relatively small DM fraction of a few percent. In this case, large DM fractions lie outside the 90\% HPD region, since they would modify the stellar structure too strongly to remain compatible with the combined NICER and GW170817 constraints after averaging over the EOS ensemble. The posterior for $\lambda$ is broad, extending over a large part of the scanned interval, which indicates that the present data do not sharply constrain the strength of the repulsive self-interaction. In contrast, the boson mass and DM fraction show clearer preferred regions. The two-dimensional marginalized posteriors also show that the preferred configurations are concentrated in the low-fraction region, with the most probable $F_\chi$ values below about $10\%$.

The regions outside the red 90\% HPD intervals in Fig.~\ref{fig:post_DM_miller} identify the least probable regions of the Maryland/Illinois posterior. From the one-dimensional marginalized distributions, the 90\% intervals are
\begin{equation}
\lambda=0.3\text{--}12.5,\ 
m_\chi=140\text{--}850~{\rm MeV},\ 
F_\chi=1\text{--}15\%.
\label{eq:maryland_90}
\end{equation}
Therefore, within this data set, both very light and very heavy bosons lie outside the $90\%$ HPD region,
$m_\chi\lesssim 140$~MeV and $m_\chi\gtrsim850$~MeV, as are large DM fractions,  $F_\chi\gtrsim15\%$. 
  For the coupling, the disfavored region is much less restrictive because the posterior remains broad and reaches close to the upper edge of the scanned range. Thus, the 90\% interval does not strongly exclude much of the coupling range, except very small values near the lower boundary of the scan, roughly $\lambda\lesssim0.3$.

Figure~\ref{fig:post_DM_watts} shows the corresponding result for the Amsterdam NICER analysis. In this case, the model-averaged posterior gives
\begin{equation}
\lambda=5.0^{+7.5}_{-4.0},\ 
m_\chi=330^{+195}_{-110}~{\rm MeV},\ 
F_\chi=12^{+4}_{-4}\% ,
\label{eq:amsterdam_60}
\end{equation}
at the 60\% HPD level. The preferred boson mass again lies in the few-hundred-MeV range, close to the value inferred from the Maryland/Illinois analysis. However, the preferred DM fraction is significantly larger. The most probable region now lies around intermediate fractions, approximately $F_\chi\sim10$--$15\%$, rather than around only a few percent. Therefore, unlike the Maryland/Illinois analysis, the Amsterdam data set favors an intermediate DM fraction instead of a very small one. This shows that the inferred amount of DM is more sensitive to the adopted NICER mass–radius analysis than the inferred boson mass.

The 90\% HPD intervals for the Amsterdam analysis are 
\begin{equation}
\lambda=0.3\text{--}12.5,\ 
m_\chi=140\text{--}775~{\rm MeV},\ 
F_\chi=3\text{--}18\%.
\label{eq:amsterdam_90}
\end{equation}
Thus, this analysis places both the low-mass and high-mass edges of the scanned boson-mass range outside the 90\% HPD, {$m_\chi\lesssim140$\,MeV and $m_\chi\gtrsim775$~MeV}. 
For the DM fraction, the Amsterdam analysis disfavors both very small and very large values, {$F_\chi\lesssim3\%$ and $F_\chi\gtrsim18\%$}. 

In this case, the low-$F_\chi$ region is less probable because the posterior is shifted toward intermediate DM fractions. As in the Maryland/Illinois case, the coupling $\lambda$ remains broadly allowed, with only very small couplings near the lower boundary of the scan lying outside the 90\% HPD region.

The two-dimensional marginalized posteriors provide a useful consistency check of these conclusions and show how the preferred regions appear in the joint parameter planes. In the $(\lambda,m_\chi)$ planes, both NICER analyses show high posterior density for boson masses of a few hundred MeV over a broad range of $\lambda$, confirming that the mass is more localized than the coupling within the adopted parameterization and priors. In the $(\lambda,F_\chi)$ planes, the coupling direction again remains broad, while the preferred fraction differs between the two data sets: the Maryland/Illinois analysis concentrates at low $F_\chi$, whereas the Amsterdam analysis shifts to intermediate $F_\chi$. The $(m_\chi,F_\chi)$ planes provide the clearest visual comparison of the two analyses. For the Maryland/Illinois case, the high-posterior-density region is concentrated around a few-hundred-MeV mass and a low DM fraction of a few percent. For the Amsterdam case, the high-posterior-density region remains at a few-hundred-MeV mass but moves to $F_\chi\sim10$--$15\%$. Thus, the two-dimensional posteriors do not introduce a qualitatively new constraint beyond the marginalized intervals, but they confirm the main posterior trends: the marginalized $m_\chi$ distribution is comparatively stable between the two NICER analyses, $F_\chi$ is more data-set dependent, and $\lambda$ remains broadly distributed under the adopted priors.  Although the 90\% intervals remain broad, both analyses show a similar posterior preference for boson masses of a few hundred MeV and consistently put outside the 90\% HPD interval the edges of the scanned mass range. In particular, $m_\chi \lesssim 140~\mathrm{MeV}$ is disfavored in both analyses, while the conservative common upper disfavored region is $m_\chi \gtrsim 850~\mathrm{MeV}$.

The comparison of the two NICER analyses provides several important conclusions. The three panels of Fig.~\ref{fig:dm_hpd_forest} make the comparison between the two analyses directly visible, and support three conclusions. First, the preferred DM fraction is more data-set dependent. The Maryland/Illinois analysis favors a low-fraction region, with $F_\chi$ of only a few percent, and disfavors large fractions. The Amsterdam analysis instead favors intermediate fractions around $10$--$15\%$ and disfavors very small fractions. Therefore, the common statement for $F_\chi$ is weaker than for $m_\chi$. Nevertheless, the extreme upper end of the scanned fraction range is disfavored in both analyses. The low-$F_\chi$ region is data-set dependent: it is favored in the Maryland/Illinois analysis but disfavored in the Amsterdam analysis.

Second, the self-interaction coupling $\lambda$ is the least sharply constrained of the three dark-sector parameters. The posterior remains broad in both NICER analyses, indicating that within the adopted $(m_\chi,\lambda)$ parameterization and priors, the marginalized posterior is more localized in $m_\chi$ and $F_\chi$ than in $\lambda$.  Only very small couplings near the lower edge of the scan lie outside the 90\% HPD region, while a broad range of $\lambda$ remains probable.

Third, the marginalized $m_\chi$ posterior is comparatively stable under the change between the two NICER data sets. Both analyses yield marginalized $m_\chi$ posteriors that peak at a few hundred MeV, while the low- and high-mass edges of the scanned range lie outside their 90\% HPD intervals. A conservative common statement is that $m_\chi\lesssim140$~MeV and $m_\chi\gtrsim850$~MeV lie outside the 90\% HPD regions of the evidence-weighted posteriors. This feature of the marginalized $m_\chi$ posterior persists for both NICER data sets after model averaging over the selected EOS ensemble.

Because the bosonic EOS depends on $m_\chi$ and $\lambda$ through the combination $\lambda/m_\chi^4$, it is useful to examine directly the corresponding derived scale
\begin{equation}
\mu_\chi \equiv \frac{m_\chi}{\lambda^{1/4}}.
\label{eq:mu_chi}
\end{equation}
We obtain the posterior of $\mu_\chi$ by transforming the evidence-weighted posterior in $(m_\chi,\lambda,F_\chi)$ according to Eq.~\eqref{eq:mu_chi} and marginalizing over $F_\chi$. For comparison, we also transform the adopted independent discrete-uniform priors on $m_\chi$ and $\lambda$, which gives the induced prior on $\mu_\chi$ shown in Fig.~\ref{fig:mu_chi_distribution}.

Figures~\ref{fig:mu_chi_distribution} and~\ref{fig:mu_chi_summary} provide a direct view of the
EOS-relevant combination of $m_\chi$ and $\lambda$, thereby complementing their separate marginalized distributions.
Both NICER analyses give nearly identical posterior modes, $\mu_\chi\simeq250$~MeV. At the 60\% HPD level we find
\begin{equation}
\begin{aligned}
\mu_\chi &= 250^{+195}_{-88}~{\rm MeV} && \text{(Amsterdam)},\\
\mu_\chi &= 251^{+313}_{-113}~{\rm MeV} && \text{(Maryland/Illinois)}.
\end{aligned}
\label{eq:mu_chi_60}
\end{equation}
The Amsterdam posterior is visibly narrower than the induced prior and is more localized than the Maryland/Illinois result. The latter remains consistent with the same preferred scale but develops a substantially longer high-$\mu_\chi$ tail. Thus, while the separate one-dimensional constraints on $m_\chi$ and $\lambda$ remain conditional on the adopted joint prior, the comparison with the induced prior shows that the data provide additional information on their EOS-relevant combination, particularly for the Amsterdam data set. We retain $(m_\chi,\lambda)$ as the primary microscopic parameters in the remainder of the discussion.

From the DM point of view, these figures are central because they identify not only the most probable regions of the bosonic-DM parameter space, represented by the 60\% HPD contours, but also the least probable regions, identified by the parts outside the 90\% HPD contours. This is particularly important because the disfavored regions are obtained after simultaneously incorporating NICER mass–radius data, the GW170817 tidal-deformability constraint, and the uncertainty of the 27 nucleonic EOSs through Bayesian evidence weighting. Thus, the analysis does not merely provide best-fit values; it also shows which combinations of $(\lambda,m_\chi,F_\chi)$ are unlikely once the observational constraints and nuclear EOS uncertainty are simultaneously taken into account.

\subsection{The nuclear EOS posterior}

\begin{figure*}[t]
\centering
\includegraphics[width=0.85\textwidth]{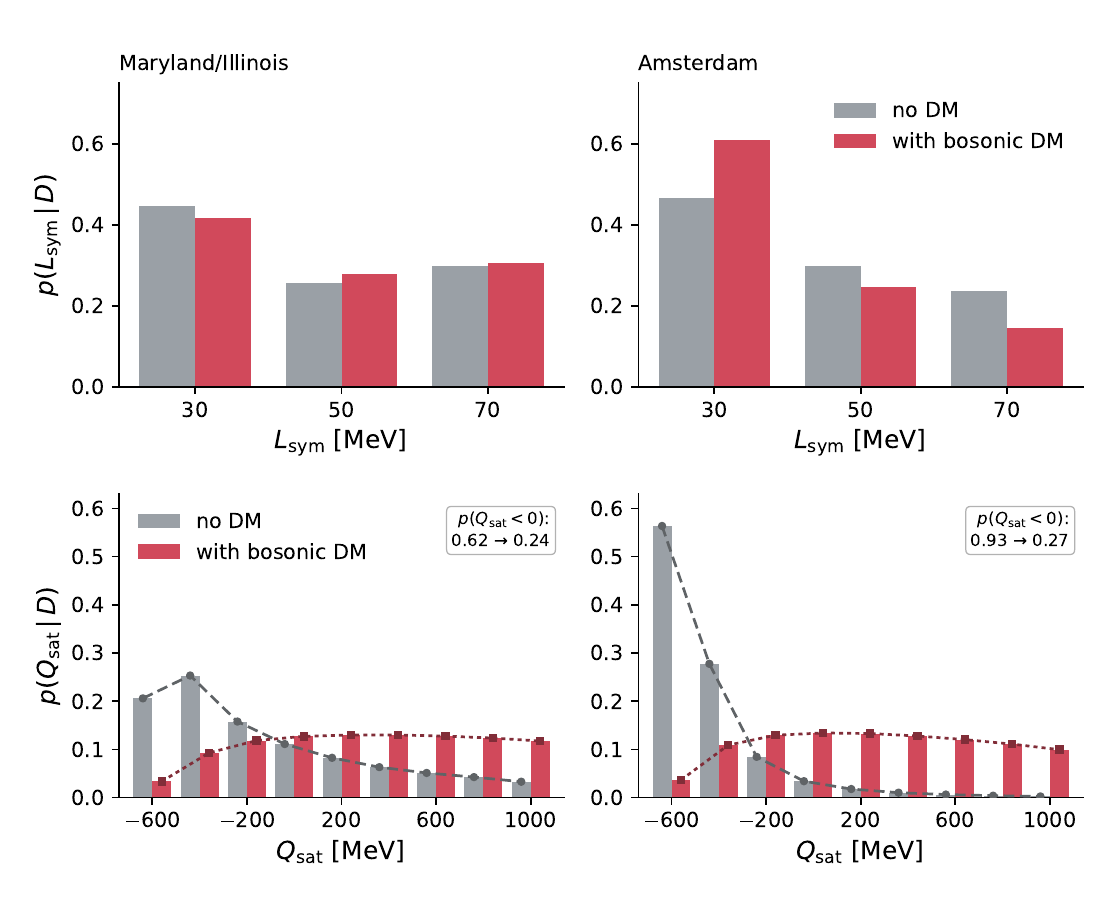}
\caption{%
Marginalized posterior probabilities of the nuclear EOS parameters $L_{\rm sym}$ (top row) and $Q_{\rm sat}$ (bottom row), obtained with the Maryland/Illinois (left column) and Amsterdam (right column) NICER mass–radius analyses, in each case combined with the GW170817 tidal-deformability constraint. Grey bars show the purely baryonic (no-DM) result and red bars the result after including self-interacting bosonic DM and marginalizing over the dark-sector parameter space. In each panel the posterior probabilities are normalized over the 27 selected CDF EOSs, so the bar heights within one series sum to unity. The light grey band in the bottom panels marks $Q_{\rm sat}<0$, i.e.\ the softer high-density sector of this CDF family. Including bosonic DM moves most of the $Q_{\rm sat}$ posterior weight from negative to non-negative values in both analyses, while the response of $L_{\rm sym}$ is weaker and data-set dependent.
}
\label{fig:LQ_marginals}
\end{figure*}

The grey bars in Fig.~\ref{fig:LQ_marginals} show the marginalized posterior distributions of the nuclear EOS parameters $L_{\rm sym}$ and $Q_{\rm sat}$ in the purely baryonic case, before introducing the bosonic-DM component. The left column corresponds to the Maryland/Illinois data set, while the right column corresponds to the Amsterdam data set. The upper panels show the marginalized posterior of $L_{\rm sym}$ and the lower panels that of $Q_{\rm sat}$. These distributions provide the reference no-DM inference for the selected CDF EOS ensemble and allow us to identify how the inferred nuclear parameters change once the dark sector is included.
The analysis is performed over the 27 nucleonic EOSs introduced in Sec.~\ref{sec:eos}, where each EOS corresponds to one pair $(L_{\rm sym},Q_{\rm sat})$. After confronting the purely baryonic stellar sequences with the observational data, each EOS receives a posterior probability. The marginalized distributions are obtained by summing these posterior probabilities over the remaining nuclear parameter. Thus, the $L_{\rm sym}$ posterior includes contributions from all sampled values of $Q_{\rm sat}$, while the $Q_{\rm sat}$ posterior includes contributions from all three sampled values of $L_{\rm sym}$.

The comparison between the two no-DM analyses shows that both NICER data sets prefer low $L_{\rm sym}$ and negative $Q_{\rm sat}$ values, but with different strengths. The Maryland/Illinois posterior is broader: it mildly favors $L_{\rm sym}=30$ MeV and peaks at the moderately negative value $Q_{\rm sat}=-400$ MeV, while still allowing a non-negligible probability over a wider range of $Q_{\rm sat}$, with $p(Q_{\rm sat}<0\,|\,D)\simeq0.62$. The Amsterdam posterior is much more concentrated: it strongly favors $L_{\rm sym}=30$ MeV and assigns most of the posterior weight to $Q_{\rm sat}=-600$ and $-400$ MeV,  giving $p(Q_{\rm sat}<0\,|\,D)\simeq0.93$. Therefore, before introducing bosonic DM, the Amsterdam analysis drives the purely baryonic EOS inference more strongly toward a soft high-density sector.

Physically, these no-DM posteriors indicate which nucleonic EOSs are preferred when the stars are assumed to be purely baryonic. Lower values of $L_{\rm sym}$ generally favor smaller radii and tidal deformabilities, while $Q_{\rm sat}$ mainly controls the high-density stiffness of the EOS. Within the present CDF family, negative $Q_{\rm sat}$ corresponds to a softer high-density behavior. Therefore, the preference for negative $Q_{\rm sat}$ indicates that, in the no-DM interpretation, the observational data favor comparatively softer high-density baryonic EOSs, with this tendency being considerably stronger for the Amsterdam data set.

Figure~\ref{fig:LQ_marginals} establishes the purely baryonic baseline for the nuclear EOS parameters. This provides the reference for assessing how the inferred $L_{\rm sym}$ and $Q_{\rm sat}$ posteriors change once bosonic DM is included. In particular, the following subsection examines whether the preference for negative $Q_{\rm sat}$, especially pronounced for the Amsterdam data set, is modified when the dark-sector degrees of freedom are allowed. This comparison provides an indication of the degeneracy between the bosonic-DM admixture and the high-density stiffness of BM.

\subsection{Impact of bosonic DM on the inferred nuclear EOS parameters}

The red bars in Fig.~\ref{fig:LQ_marginals} show the marginalized posterior distributions of the nuclear EOS parameters $L_{\rm sym}$ and $Q_{\rm sat}$ after including bosonic DM, overlaid directly on the purely baryonic result (grey) for the same data set. In the present case, the posterior probability assigned to each nucleonic EOS is obtained after allowing for the bosonic-DM component and marginalizing over the dark-sector parameter space. The overlay therefore shows directly how the preferred region of the nuclear EOS ensemble changes when the star is allowed to contain self-interacting bosonic DM.

For the Maryland/Illinois analysis, including bosonic DM makes the
$L_{\rm sym}$ posterior more balanced among the three sampled values
and shifts the $Q_{\rm sat}$ distribution from its no-DM peak at
$Q_{\rm sat}=-400~\mathrm{MeV}$ toward a broad maximum around
$Q_{\rm sat}=200~\mathrm{MeV}$. For the Amsterdam analysis, the
preference for $L_{\rm sym}=30~\mathrm{MeV}$ remains, while the
$Q_{\rm sat}$ posterior changes much more strongly, moving from the
very negative no-DM region toward a broad maximum around
$Q_{\rm sat}=0~\mathrm{MeV}$.

Overall, Fig.~\ref{fig:LQ_marginals} summarizes this change in the nuclear EOS posterior. The main effect of bosonic DM is a redistribution of probability in $Q_{\rm sat}$: the no-DM posterior favors negative values, while the DM-admixed posterior shifts most of the weight to $Q_{\rm sat}\ge0$ in both NICER analyses, with $p(Q_{\rm sat}<0\,|\,D)$ dropping from $0.62$ to $0.24$ for the Maryland/Illinois set and from $0.93$ to $0.27$ for the Amsterdam set. The response of $L_{\rm sym}$ is weaker and more data-set dependent, becoming more balanced for the Maryland/Illinois set but remaining concentrated at $L_{\rm sym}=30$ MeV for the Amsterdam set.

This shift has a simple physical interpretation. In a purely baryonic model, compact mass–radius measurements are accommodated by selecting a softer high-density EOS, corresponding here to negative $Q_{\rm sat}$. When bosonic DM is allowed, part of the required reduction of the visible radius can be produced by the dark component itself. The baryonic EOS is then no longer forced to be as soft as in the no-DM interpretation, and the posterior shifts toward intermediate and positive $Q_{\rm sat}$ values.
This indicates a model-dependent trade-off between the bosonic-DM admixture and the high-density stiffness of baryonic
matter.

\subsection{Single-constraint Bayesian comparison between the DM-admixed and no-DM model families}
\label{subsec:evidence_weighted_dm_nodm}

The previous subsections discussed the evidence-weighted posterior constraints on the bosonic-DM parameters and the corresponding changes in the inferred nuclear EOS parameters. Here we ask a complementary model-comparison question: whether the DM-admixed model family is globally preferred over the purely nucleonic no-DM model family once the same 27 baryonic EOSs are taken into account.

For the DM-admixed case, the Bayesian evidence is first computed for each fixed baryonic EOS, specified by the pair $(L_{\rm sym},Q_{\rm sat})$, after marginalizing over the dark-sector parameters $(\lambda,m_\chi,F_\chi)$. We denote this fixed-EOS evidence by
\begin{equation}
Z_{\rm DM}^{X}(L_{\rm sym},Q_{\rm sat}),
\end{equation}
 {where $X$ labels the adopted NICER data set, namely the Amsterdam or Maryland/Illinois analysis.} Assuming equal prior probability for the 27 baryonic EOSs, the global DM evidence is
\begin{equation}
Z_{{\rm DM,global}}^{X}
=
\frac{1}{27}
\sum_{L_{\rm sym},Q_{\rm sat}}
Z_{\rm DM}^{X}(L_{\rm sym},Q_{\rm sat}).
\label{eq:z_dm_global}
\end{equation}
The corresponding no-DM evidence, $Z_{\rm noDM}^{X}$, is obtained from the purely nucleonic Bayesian analysis over the same EOS ensemble. We then define the evidence ratio
\begin{equation}
B_{\rm DM/noDM}^{X}
=
\frac{Z_{{\rm DM,global}}^{X}}{Z_{\rm noDM}^{X}}.
\label{eq:bayes_factor_dm_nodm}
\end{equation}
For equal prior odds between the DM-admixed and no-DM model families, this quantity is the Bayes factor in favor of the DM-admixed interpretation.  We use the KDE reconstructions of the published observational posteriors as effective likelihood densities and apply the same curve-overlap prescription uniformly to all EOS and dark-sector models. The posterior constraints and evidence ratios quoted below should therefore be understood as conditional on this likelihood construction.

The resulting evidences are listed in Table~\ref{tab:global_evidence_dm_nodm}. For the Amsterdam set, the evidence ratio is $B_{\rm DM/noDM}\simeq2.19$, which indicates a weak Bayesian preference for the DM-admixed model under the adopted priors, but does not constitute robust evidence for a dark component. For the Maryland/Illinois set $B_{\rm DM/noDM}<1$, i.e. no global preference. The evidence-level comparison is therefore data-set dependent: the Amsterdam set shows a slight preference for the inclusion of bosonic DM, whereas the Maryland/Illinois set does not favor it.
We caution that since the upper bounds on $\lambda$  and $m_\chi$ were chosen heuristically rather than derived from an independent physical argument, the specific values 2.19 and 0.82 should be interpreted only as indicating the same order of (weak, data-set-dependent) preference rather than as precise numbers.
\begin{table}[t]
\centering
\caption{Global Bayesian evidence comparison between the DM-admixed and no-DM model families. The global DM evidence is obtained by averaging the fixed-EOS DM evidences over the 27 baryonic EOSs using Eq.~\eqref{eq:z_dm_global}.}
\label{tab:global_evidence_dm_nodm}
\begin{ruledtabular}
\begin{tabular}{lccc}
Data set
& $Z_{{\rm DM,global}}$
& $Z_{\rm noDM}$
& $B_{\rm DM/noDM}$ \\
Amsterdam set
& $1.33\times10^{-6}$ 
& $6.09\times10^{-7}$ 
& $2.19$ \\
Maryland/Illinois set 
& $1.03\times10^{-6}$ 
& $1.25\times10^{-6}$ 
& $0.822$ \\
\end{tabular}
\end{ruledtabular}
\end{table}

To identify how the individual observational constraints contribute to the model comparison, we additionally compute a Bayes factor for each constraint separately. For a given observation $E_k$, we evaluate the evidence directly from the adopted priors. For the DM-admixed model,
\begin{equation}
Z_{{\rm DM}|E_k}
=
\frac{1}{27}\sum_{e=1}^{27}\sum_{\theta}
\pi(\theta)\,\mathcal{L}^{\rm DM}_{k}(E_k|e,\theta),
\label{eq:single_Z_DM}
\end{equation}
where $e$ labels the baryonic EOS, $\theta=(m_\chi,\lambda,F_\chi)$, and $\pi(\theta)=1/24200$ is the discrete-uniform prior over the sampled DM parameter space. For the no-DM model,
\begin{equation}
Z_{{\rm noDM}|E_k}
=
\frac{1}{27}\sum_{e=1}^{27}
\mathcal{L}^{\rm noDM}_{k}(E_k|e).
\label{eq:single_Z_noDM}
\end{equation}
The corresponding single-constraint Bayes factor is
\begin{equation}
B_{\rm DM/noDM}^{k}
=
\frac{Z_{{\rm DM}|E_k}}{Z_{{\rm noDM}|E_k}} .
\label{eq:single_BF}
\end{equation}
Thus, $B_{\rm DM/noDM}^{k}>1$ indicates that the observation, considered independently, favors the DM-admixed model family, whereas $B_{\rm DM/noDM}^{k}<1$ favors the no-DM family.

\begin{figure*}[t]
    \centering
    \includegraphics[width=1.5\columnwidth]{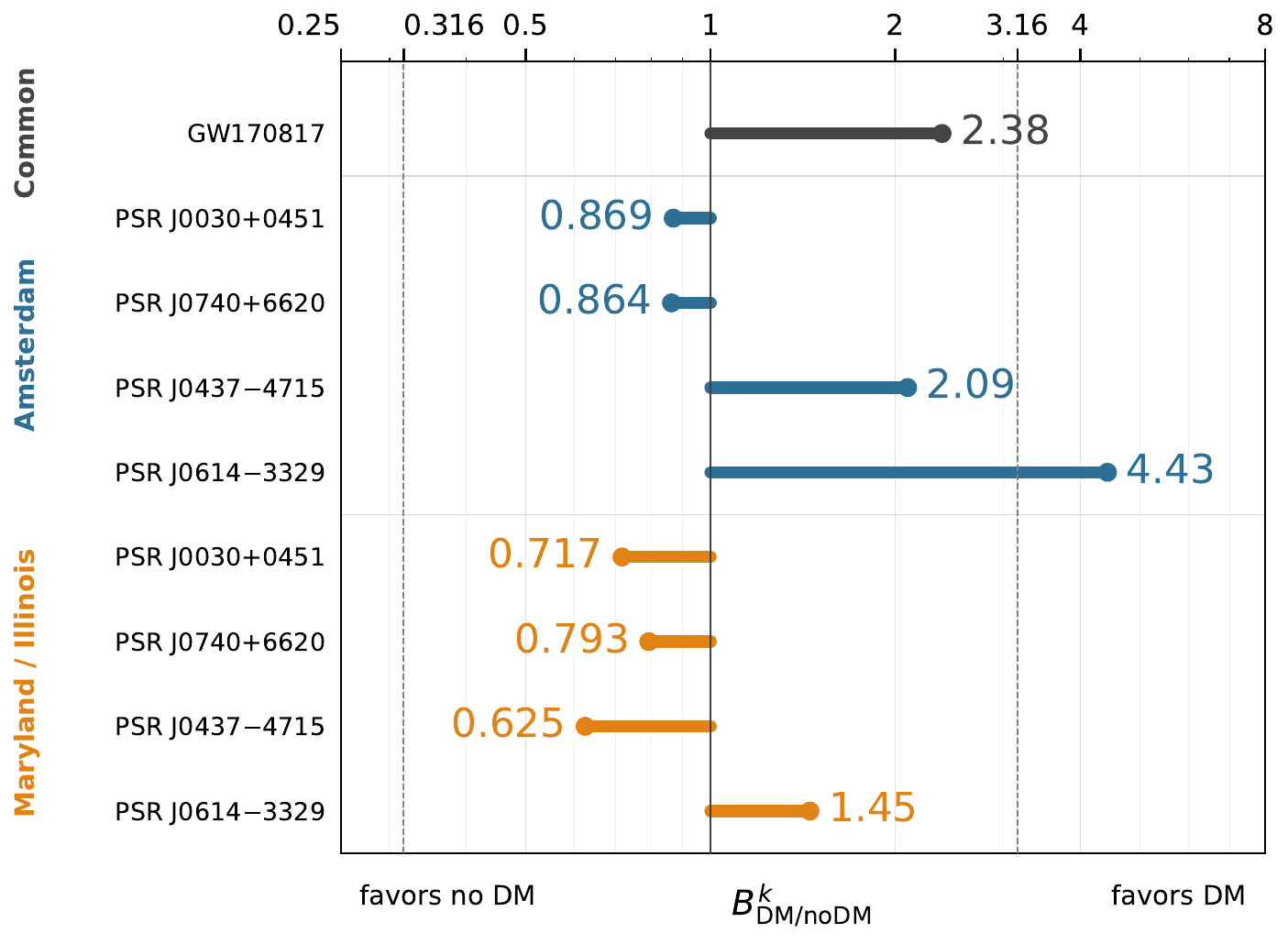}
    \caption{
    Single-constraint Bayes factors $B_{\rm DM/noDM}^{k}$ comparing the bosonic-DM-admixed and no-DM model families. Each value is obtained using only the indicated observational constraint, while marginalizing over the 27 baryonic EOSs and, for the DM-admixed model, over the sampled $(m_\chi,\lambda,F_\chi)$ parameter space with the adopted discrete-uniform priors. Values larger than unity favor the DM-admixed model, whereas values smaller than unity favor the no-DM model. The vertical dashed lines at $B_{\rm DM/noDM}^{k}\simeq0.316$ and $3.16$
($10^{-1/2}$ and $10^{1/2}$, respectively) indicate the boundary between
weak and substantial evidence in favor of the corresponding model family
according to the Jeffreys scale.}
    \label{fig:single_constraint_bf}
\end{figure*}

Figure~\ref{fig:single_constraint_bf} shows that the single-constraint support for bosonic DM is neither uniform across observations nor robust with respect to the adopted NICER analysis. Following the Jeffreys evidence scale \cite{Jeffreys1961}, we use $B=10^{1/2}$ and $B=10^{-1/2}$ as the conventional boundaries for substantial evidence in favor of the DM-admixed and no-DM model families, respectively. Among the individual constraints, only PSR J0614-3329 from the Amsterdam set yields substantial evidence for the DM-admixed family, with $B^{k}_{\rm DM/noDM}=4.43>10^{1/2}$, whereas GW170817, as well as PSR J0437-4715 from the Amsterdam analysis, provide only weak support, with Bayes factors of 2.38 and 2.09, respectively. PSR J0740+6620 and PSR J0030+0451 are essentially non-discriminating for both NICER sets. The PSR J0437-4715 constraint from the Maryland/Illinois analysis yields Bayes factors below unity, in contrast to the Amsterdam analysis. However, PSR J0614-3329 shows a trend similar to its Amsterdam counterpart, yielding Bayes factors above unity, while the preference remains very weak. Notably, none of the constraints reaches the threshold $10^{-1/2}$ for substantial evidence in favor of the no-DM family. Particularly informative is the different behavior for the same pulsars between the Amsterdam and Maryland/Illinois analyses, indicating that the inferred preference is sensitive to differences in the adopted mass–radius posteriors. The tendency of the more compact constraints to favor the DM-admixed family is physically consistent with the compactifying effect of core-like DM configurations, which can reduce the visible baryonic radius and partially mimic a softer baryonic EOS.

 Figure~\ref{fig:single_constraint_bf} therefore shows the preference associated with each observation considered separately, whereas Table~\ref{tab:global_evidence_dm_nodm} gives the model comparison for the complete multimessenger data combinations. The two comparisons are consistent, while the single-constraint results help identify which observations are associated with the different preferences of the two data combinations.

The Bayesian evidence does not imply a universal preference for bosonic DM across all observational choices. The Amsterdam data combination provides weak model-level support for the DM-admixed scenario, whereas the Maryland/Illinois combination does not favor it. The single-constraint Bayes factors also show that the preference differs among the individual observations. Taken together, these results show that the current support for bosonic DM depends on the adopted observational data set and is not robust across the different analyses.

\section{Discussion and Conclusions}
\label{sec:conclusions}

We have investigated NSs admixed with repulsively
self-interacting bosonic dark matter DM within a two-fluid framework.
The baryonic sector was represented by 27 DDME2-based CDF EOSs
spanning systematic variations of $L_{\rm sym}$ and $Q_{\rm sat}$,
while the dark sector was characterized by $(m_\chi,\lambda,F_\chi)$.
For each baryonic EOS, the stellar sequences were confronted with
NICER mass--radius information from the Amsterdam and
Maryland/Illinois analyses and with the GW170817 tidal-deformability
constraint. The resulting fixed-EOS posteriors were combined by
Bayesian evidence weighting, thereby propagating the uncertainty
within the selected CDF ensemble into the inferred dark-sector
constraints.

Within the adopted parameterization and priors, the marginalized
$m_\chi$ posterior peaks at masses of a few hundred MeV for both
NICER data sets, whereas $\lambda$ remains only weakly localized.
The inferred DM fraction is more data-set dependent:
the Maryland/Illinois analysis favors a fraction of a few percent,
while the Amsterdam analysis favors $F_\chi\sim10$--$15\%$.
The corresponding credible intervals for $m_\chi$ remain broad. Since the adopted bosonic EOS depends on $m_\chi$ and $\lambda$ through the combination $\lambda/m_\chi^4$, their separate marginalized constraints are conditional on the adopted parameterization and discrete-uniform priors. As a complementary representation, the derived scale $\mu_\chi=m_\chi/\lambda^{1/4}$ peaks near $250$~MeV for both NICER data sets, with the Amsterdam posterior more strongly localized relative to the induced prior than the Maryland/Illinois result. Structurally, lighter bosons and stronger repulsive self-interactions favor more pressure supported dark distributions and extended halos, whereas heavier bosons and weaker self-interactions favor compact dark cores.

An important result concerns the interplay between the dark component
and the high-density baryonic EOS. In the purely baryonic analysis,
both NICER data sets favor negative $Q_{\rm sat}$, particularly the
Amsterdam set. Once bosonic DM is included and marginalized over,
the posterior weight shifts toward intermediate and positive
$Q_{\rm sat}$ values: the probability for $Q_{\rm sat}<0$ decreases
from $0.62$ to $0.24$ for the Maryland/Illinois data and from $0.93$
to $0.27$ for the Amsterdam data. The response of $L_{\rm sym}$ is
considerably weaker. This behavior indicates a model-dependent
trade-off between the compactifying effect of the dark component and
the high-density stiffness of BM. Compact stellar
configurations that require a relatively soft baryonic EOS in the
no-DM interpretation can also be accommodated by somewhat stiffer
EOSs when part of the radius reduction is produced by a dark
component.

The shift of the $Q_{\rm sat}$ posterior toward intermediate and positive
values when bosonic DM is included should not be interpreted as a unique
signature of a dark component. Within the restricted nucleonic CDF
ensemble considered here, additional high-density degrees of freedom or
phase structure that produce comparable stellar compactification---for
example, hyperonization, deconfined quark matter, or a sufficiently strong
phase transition---could reduce the need to reproduce compact stars through
a negative $Q_{\rm sat}$ and thereby induce a qualitatively similar
compensating shift toward a stiffer underlying nucleonic sector
\cite{Sedrakian:2023, Baym:2018}. The redistribution of $Q_{\rm sat}$ in
Fig.~\ref{fig:LQ_marginals} should therefore be viewed more generally as evidence of
a degeneracy between the assumed high-density microphysics and the
inferred stiffness of the nucleonic EOS, rather than as a unique diagnostic
of bosonic DM.

The Bayesian model comparison remains inconclusive. Within the
adopted priors and likelihood construction, the Amsterdam data yield
only a weak preference for the DM-admixed model family,
$B_{\rm DM/noDM}\simeq2.2$, whereas the Maryland/Illinois data give
$B_{\rm DM/noDM}\simeq0.82$ and therefore do not favor the inclusion
of DM. The current observations thus provide no robust evidence for
a bosonic dark component in NSs. The quantitative
constraints are also conditional on the restricted DDME2-based EOS
ensemble and on the assumption of a common effective $F_\chi$ for
all sources. In reality, the dark fraction may depend on formation
history, environment, capture or conversion mechanisms, and stellar
properties. In addition, standard NICER pulse-profile analyses do not
include an extended gravitating dark halo, although the most extended
halo configurations considered here receive very small weight from
the GW170817 tidal-deformability constraint.

Future observations that combine more precise masses and radii with
gravitational-wave, thermal, neutrino, and pulse-profile information
will be important for separating dark-sector effects from conventional
dense-matter physics. In particular, source-dependent dark fractions,
broader baryonic EOS ensembles, and a consistent treatment of
halo-modified pulse profiles would provide natural extensions of the
present analysis. Thermal and compositional observables may offer
additional discriminatory power, since dark cores and exotic baryonic
degrees of freedom need not affect the thermal structure in the same
way~\cite{Issifu:2025jac}. Such complementary probes will be required
before robust particle-physics constraints on bosonic DM can
be extracted from NS observations.

\begin{acknowledgments}
D.~R.~K. acknowledges support from the Polish National Science Centre (NCN) under Grant No.~2023/51/B/ST9/02798 and from the SONATINA~7 program under Grant No.~2023/48/C/ST2/00297. Part of this work was carried out during D.~R.~K.'s stay at the Few-body Systems in Physics Laboratory, RIKEN Nishina Center for Accelerator-Based Science, Japan. He gratefully acknowledges the hospitality of the laboratory, in particular Prof.~Emiko Hiyama and Prof.~Makoto Oka, during the completion of this work.

A.~A. acknowledges support from project No.~2021/43/P/ST2/03319, co-funded by the National Science Centre and the European Union's Horizon 2020 research and innovation programme under the Marie Sk\l{}odowska-Curie Grant Agreement No.~945339.

A.~S. acknowledges support from the Polish National Science Centre (NCN) under Grant No.~2023/51/B/ST9/02798 and partial support from Collaborative Research Grant No.~24RL-1C010, provided by the Higher Education and Science Committee (HESC) of the Republic of Armenia through the ``Remote Laboratory'' program.
\end{acknowledgments}

\bibliographystyle{apsrev4-2}
\bibliography{sample631}

\end{document}